\documentclass[iop]{emulateapj}
\usepackage{color}
\usepackage{natbib}
\usepackage{amsmath,amssymb}

\shorttitle{NGC346 Turn-Ons}
\shortauthors{Cignoni et al.}

\begin{document}

\title{Pre-Main sequence Turn-On as a chronometer for young clusters:
  NGC346 as a benchmark\altaffilmark{1}}

\author{M. Cignoni\altaffilmark{2,3}, M. Tosi\altaffilmark{3},
  E. Sabbi\altaffilmark{4}, A.  Nota\altaffilmark{4,5},
  S. Degl'Innocenti\altaffilmark{6,7}, P.G. Prada
  Moroni\altaffilmark{6,7}, J. S.  Gallagher\altaffilmark{8}}

\altaffiltext{1}{Based on observations with the NASA/ESA Hubble Space
  Telescope, obtained at the Space Telescope Science Institute, which
  is operated by the Association of Universities for Research in
  Astronomy (AURA), Inc., under NASA contract NAS5-26555. These
  observations are associated with program GO10248.}

\altaffiltext{2}{Dipartimento di Astronomia, Universit\`a degli Studi di
  Bologna, via Ranzani 1, I-40127 Bologna, Italy}

\altaffiltext{3}{Istituto Nazionale di Astrofisica, Osservatorio
  Astronomico di Bologna, Via Ranzani 1, I-40127 Bologna, Italy}
\altaffiltext{4}{Space Telescope Science Institute, 3700 San Martin
  Drive, Baltimore, USA} \altaffiltext{5}{European Space Agency,
  Research and Scientific Support Department, Baltimore, USA}
\altaffiltext{6}{Dipartimento di Fisica ``Enrico Fermi'', Universit\`a
  di Pisa, largo Pontecorvo 3, Pisa I-56127, Italy}
\altaffiltext{7}{INFN - Sezione di Pisa, largo Pontecorvo 3, Pisa
  I-56127, Italy} \altaffiltext{8}{Department of Astronomy, 475
  N. Charter St., Madison, WI 53706 USA}

\begin{abstract}

We present a novel approach to derive the age of very young star
clusters, by using the Turn-On (TOn). The TOn is the point in the
color-magnitude diagram (CMD) where the pre-main sequence (PMS) joins
the main sequence (MS). In the MS luminosity function (LF) of the
cluster, the TOn is identified as a peak followed by a dip. We propose
that by combining the CMD analysis with the monitoring of the spatial
distribution of MS stars it is possible to reliably identify the TOn
in extragalactic star forming regions. Compared to alternative
methods, this technique is complementary to the turn-off dating and
avoids the systematic biases affecting the PMS phase. We describe the
method and its uncertainties, and apply it to the star forming region
NGC346, which has been extensively imaged with the Hubble Space
Telescope (HST). This study extends the LF approach in crowded
extragalactic regions and opens the way for future studies with
HST/WFC3, JWST and from the ground with adaptive optics.
\end{abstract}


\section{Introduction}

The MS turn-off is the most reliable feature for age-dating a star
cluster.  Nevertheless for very young clusters, with ages of tenths to
few Myr, the identification of the turn-off is usually hampered by the
paucity of massive stars.

We propose to take a different point of view, focusing on the TOn, the
CMD locus where the PMS joins the MS. Although the importance of the
TOn has been already emphasized in several papers (e.g.
\citealt{Stauffer80}, \citealt{Belikov98}, \citealt{Baume03},
\citealt{Naylor09}), its application to dating extragalactic star
forming regions is a new proposition.

In analogy with the turn-off, the TOn properties are directly related
to the age of the stellar population, but with evolutionary times much
shorter than the corresponding MS times. In fact, from simple stellar
evolution arguments, the age of a cluster is equal to the time spent
in the PMS phase by its most massive star still in the PMS phase. By
definition, this star is at the TOn. Hence, when the intrinsic
luminosity of the TOn is detected, it is straightforward to associate
it to the age of the cluster.

In the first part of this letter we describe how the intrinsic
properties of the TOn can be used as a clock. In the second part we
present a new method to apply TOn related properties to date
extragalactic systems. Finally we apply this method to the largest
extragalactic star forming region, NGC346, in the Small Magellanic
Cloud (SMC).

\section{The PMS Turn-On}

\begin{figure*}[t!]
\centering\includegraphics[width=6cm]{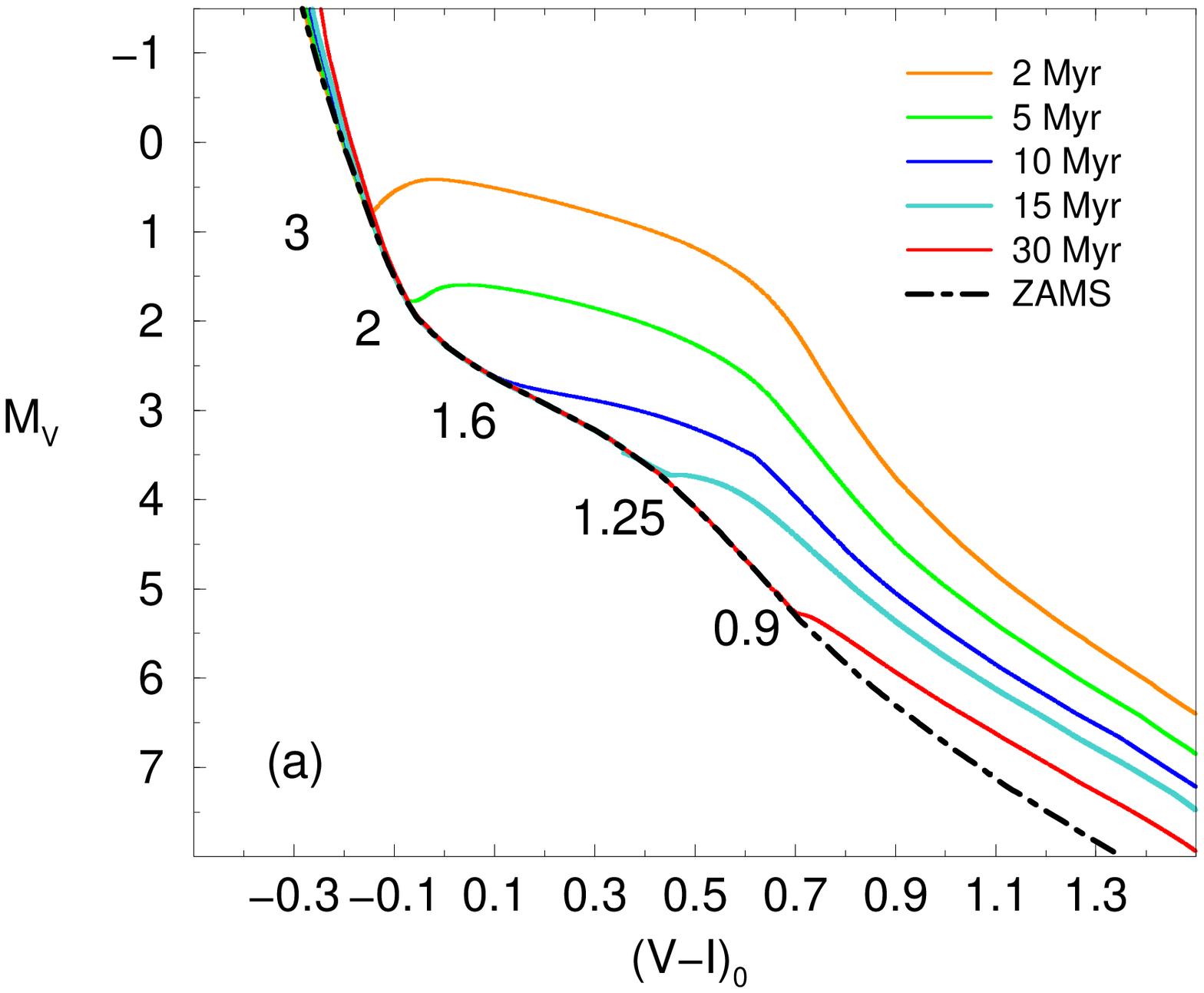} 
\centering\includegraphics[width=6cm]{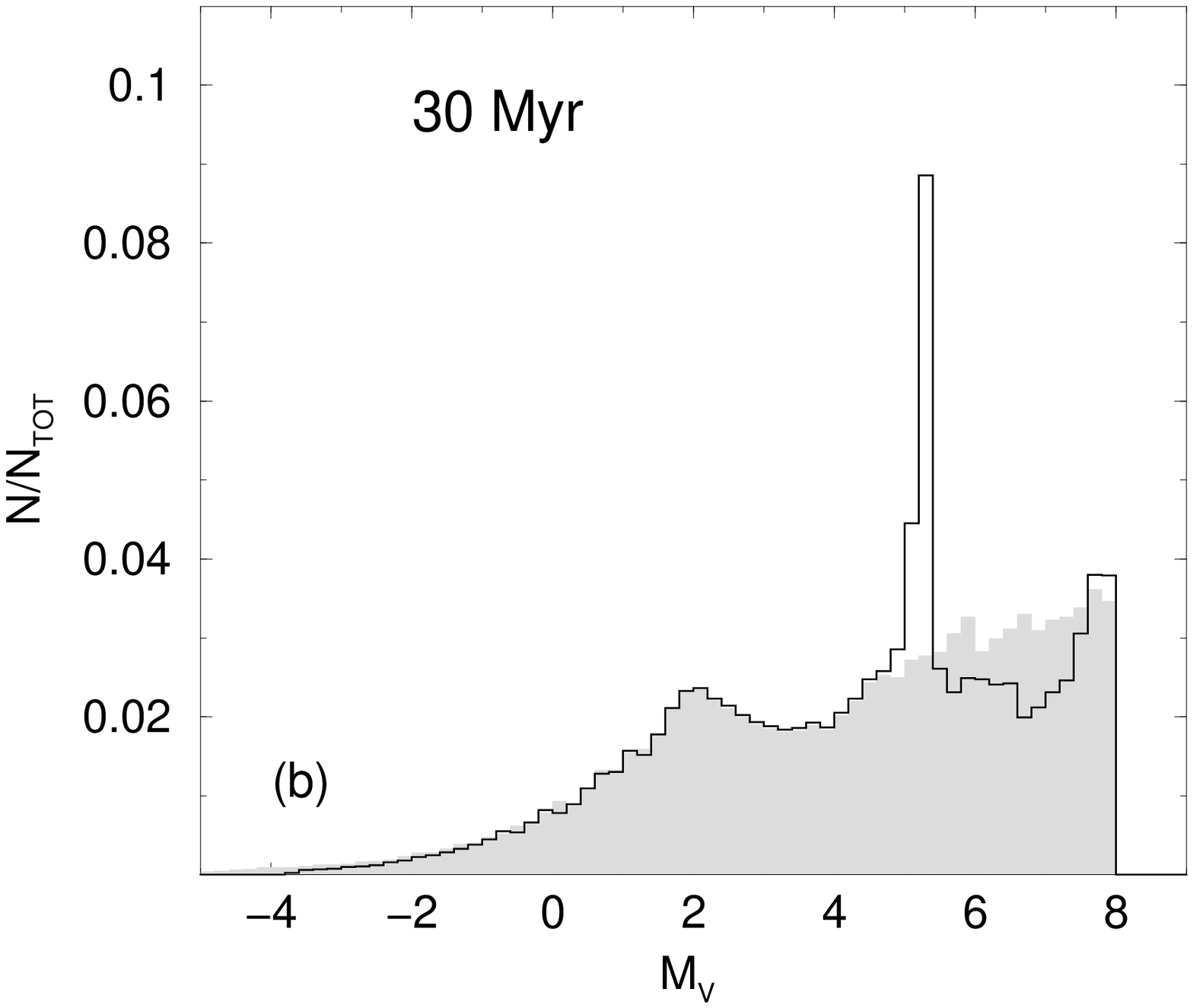}\\
\centering\includegraphics[width=6cm]{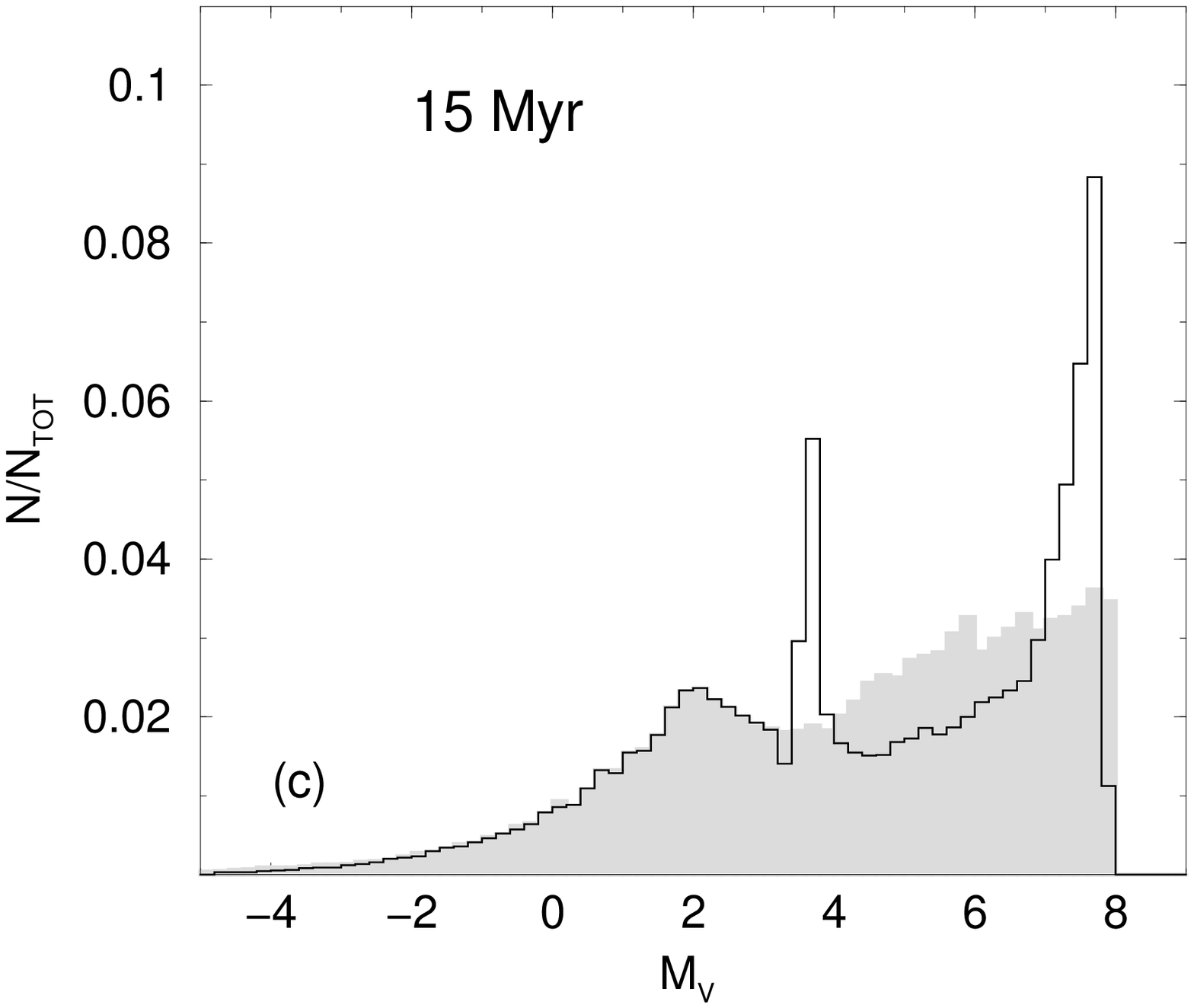}
\centering\includegraphics[width=6cm]{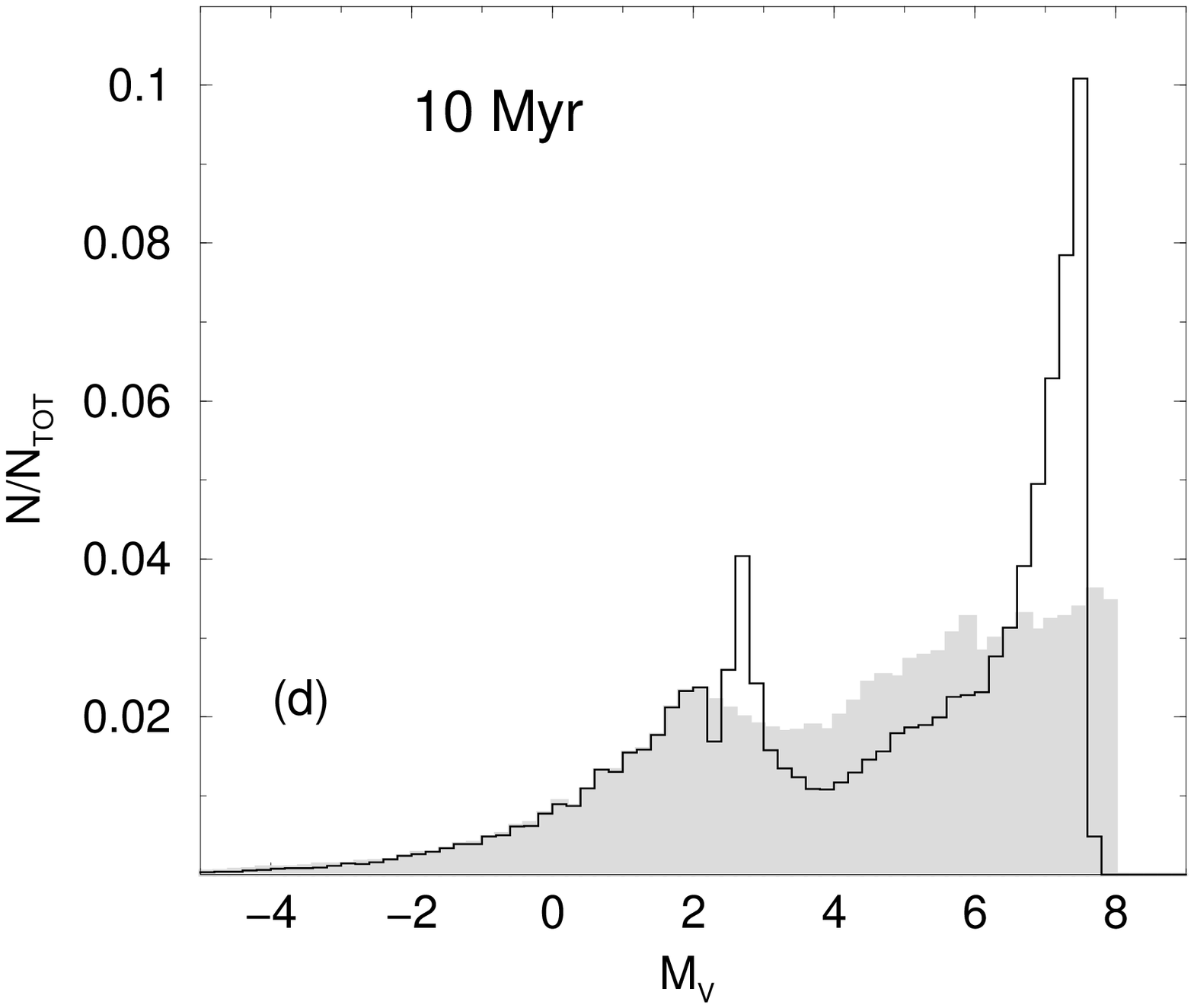}\\ 
\centering\includegraphics[width=6cm]{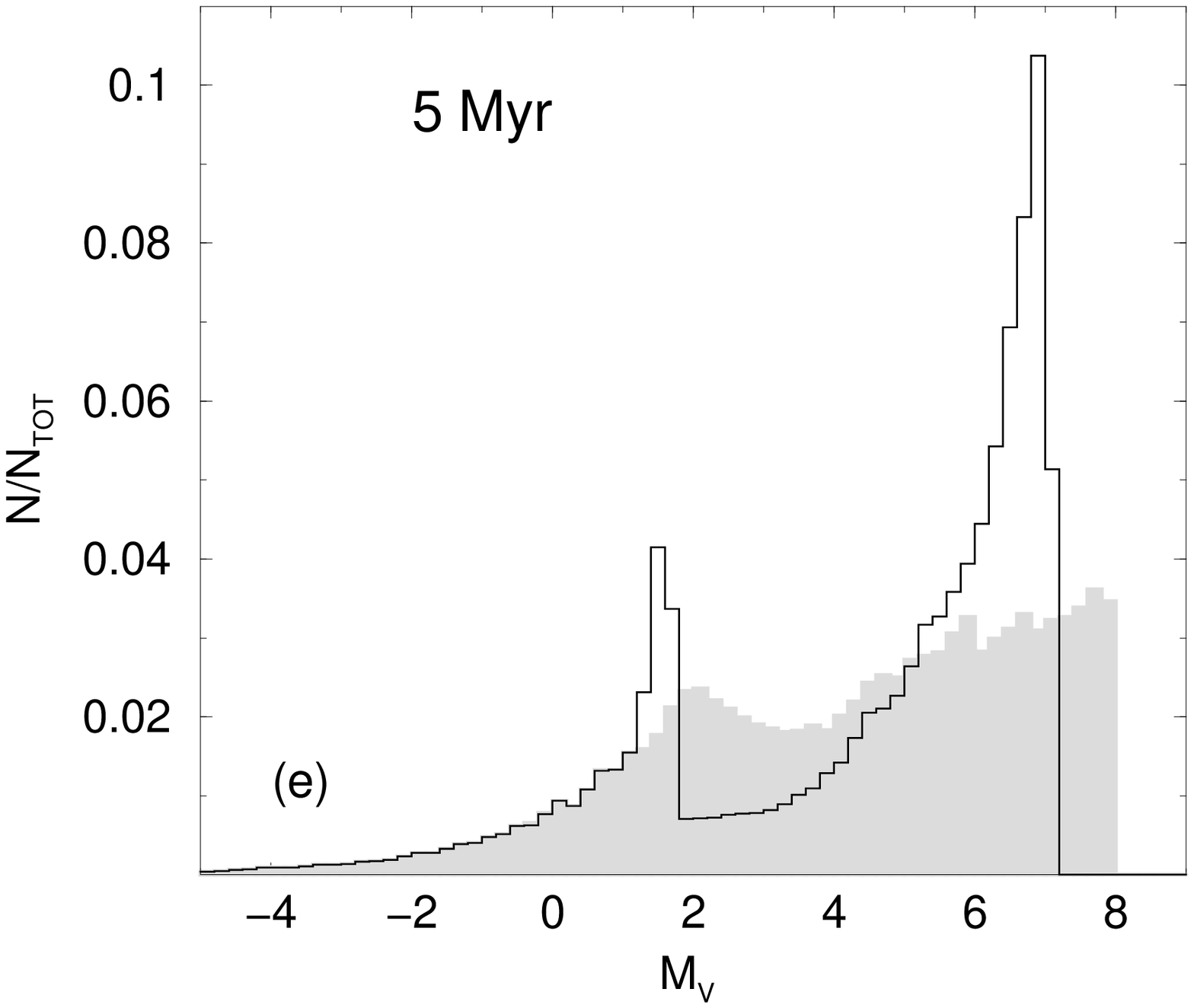}
\centering\includegraphics[width=6cm]{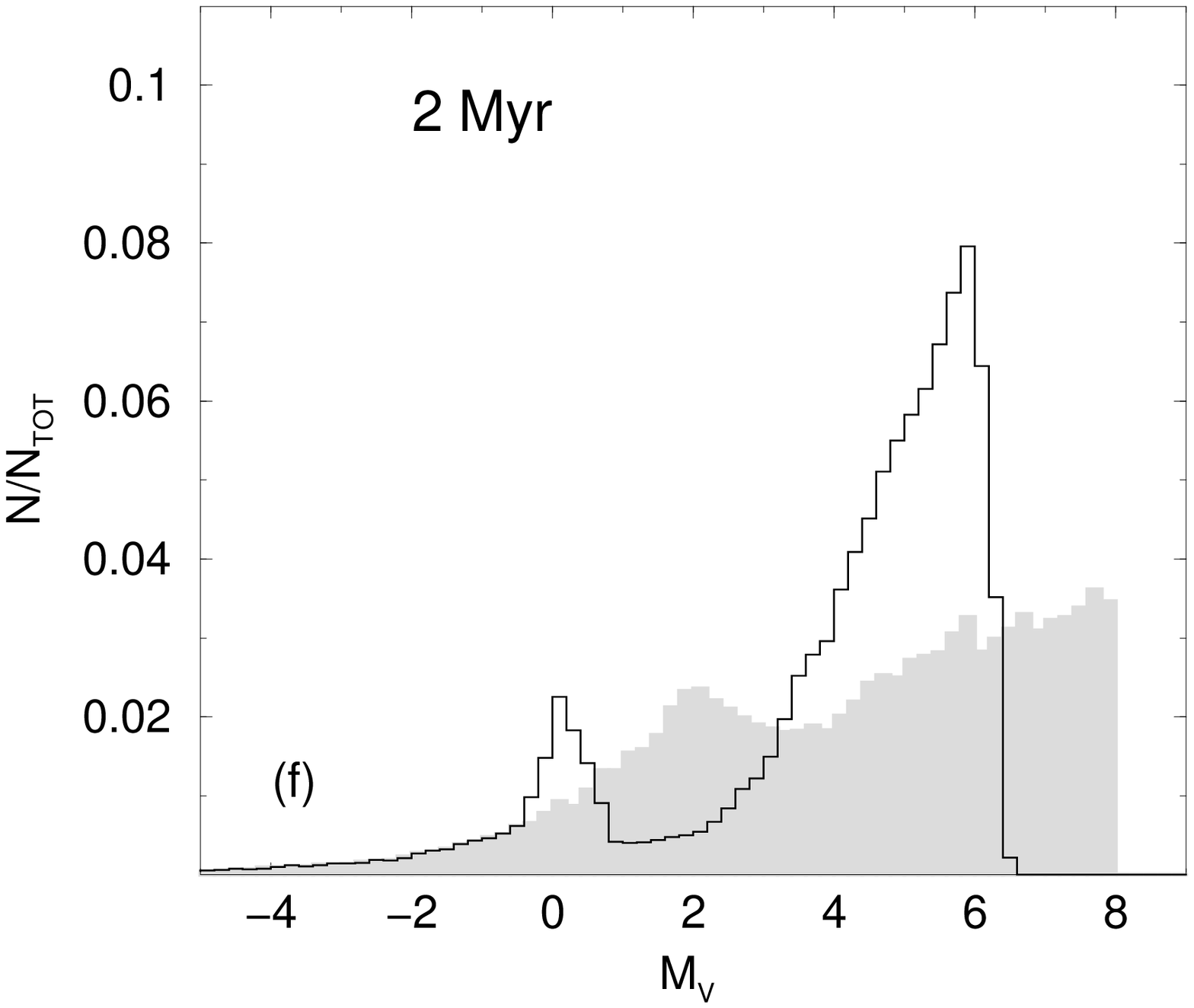}
\caption{Panel (a): combined PMS and MS isochrones for the labeled
  ages. The TOn is at the discontinuity where the PMS joins the MS
  (e.g. at $(V-I)_0=-0.15$ and $M_{V}=0.85$ for the 2 Myr case). The
  numbers on the left give the mass of the TOn star of each
  isochrone. Panels (b) to (f): normalized LFs for synthetic SSPs of
  the labeled age (see text for details).}
\label{peaks} 
\end{figure*}

The potential strength of the TOn is apparent from the morphology of
isochrones taking both the PMS and MS phases into
account. Fig.\ref{peaks}(a) shows 5 isochrones with metallicity
$Z=0.004$, obtained combining the Pisa PMS tracks
(\citealt{Cignoni09}) with the Padua evolutionary tracks
(\citealt{Fagotto94}) to cover the entire mass range
$0.45-120\,M_{\odot}$.  For all ages younger than 15 Myr the isochrone
portion just before the zero age MS (ZAMS) has a hook and then is
significantly flatter than the ZAMS. The TOn is at the vertex of the
hook, quite easy to recognize.

To test if/how the TOn can be useful as a cosmo-chronometer, we
simulated synthetic simple stellar populations (SSPs), using the
tracks quoted above. As an example, we describe the case with
metallicity $Z=0.004$, burst duration 1 Myr, Salpeter's Initial Mass
Function (IMF), and no binaries. To guarantee statistically
significant tests, extensive theoretical simulations have been
performed with a number of synthetic stars up to $1\times 10^6$.

Panels (b)-(f) in Fig.\ref{peaks} show the luminosity functions (LFs),
binned in 0.2 magnitude bins and scaled to the total number of stars,
of synthetic SSPs (open histograms) corresponding to the isochrones of
panel (a). A reference synthetic zero age population, that is
artificially built without\footnote{i.e. stars are born directly on
  the ZAMS.} PMS stars (grey filled histogram), is also shown. In the
LF without PMS, the only valuable feature is a mild peak around
$M_{V}\approx 2$, consequence of an inflection in the derivative of
the mass-$M_{V}$ relation in MS stellar models around
$m=2\,M_{\odot}$. By contrast, when the evolution starts from the PMS
phase, the corresponding LFs develop an additional strong peak
followed by a dip. Peak and dip reflect respectively the steep
dependence of stellar mass on $M_{V}$ near the TOn (see also the
discussion in \citealt{piskunov}) and the following flattening below
the TOn (caused by the short evolutionary timescale of the PMS phase
compared to the MS). After the dip, the shape of the LF mimics the
IMF.

The LFs in Fig.\ref{peaks} indicate the importance of both features,
peak and dip, to infer the cluster age: the older the age, the fainter
is the LF TOn peak and the corresponding dip. On the other hand, for
the explored range of ages the magnitude of the MS peak is fairly
constant ($M_{V}\approx 2$), being it locked to the (much longer) MS
evolutionary times. Through a polynomial fit to the models, theory
provides a useful relation between age ($\tau$) of the SSP and the
magnitude $M_{V}$ of the TOn in the range 2-100 Myr:
\begin{eqnarray} 
\displaystyle  
\tau (Myr)=&\sum_{j=0,5} a_{j}\times (M_{V})^j& \label{eq1}
\end{eqnarray}
In turn:
\begin{eqnarray} 
\displaystyle
M_{V}=&\sum_{j=0,5} b_{j}\times \{\log  [\tau(Myr)]\}^j&\label{eq2}
\end{eqnarray}

\begin{table}[h!]
\begin{center}
\caption{Coefficients for Eq. 1 and Eq. 2.}
\label{tab}
\vspace{0.2cm}
\begin{tabular}{c|c|c} \hline
\multicolumn{1}{c|}{j}&\multicolumn{1}{|c}{$a_{j}$}&\multicolumn{1}{|c}{$b_{j}$}\\ \hline
0&8.144&-2.595\\ \hline 
1&-17.620&22.021\\ \hline
2&16.120&-48.826\\ \hline
3&-5.005&51.418\\ \hline
4&0.6908&-23.420\\ \hline
5&-0.03218&3.904\\ \hline
\end{tabular}
\end{center}
\end{table}

Once the bin width of the LF is chosen, Eq. (\ref{eq1}) (whose
coefficients are given in Table \ref{tab}) gives also the intrinsic
uncertainty on the age. Assuming 0.2 mag wide bins, the minimum
uncertainty is about 0.6 Myr at 3 Myr, 1.3 Myr at 20 Myr, 2.5 Myr at
30 Myr and 6 Myr at 50 Myr. The TOn formula is deliberately limited to
populations older than 2 Myr, since for younger ages the current PMS
models are still extremely uncertain (see \citealt[][ for a
  discussion]{baraffe02}).

\begin{figure}[]
\centering\includegraphics[width=7cm]{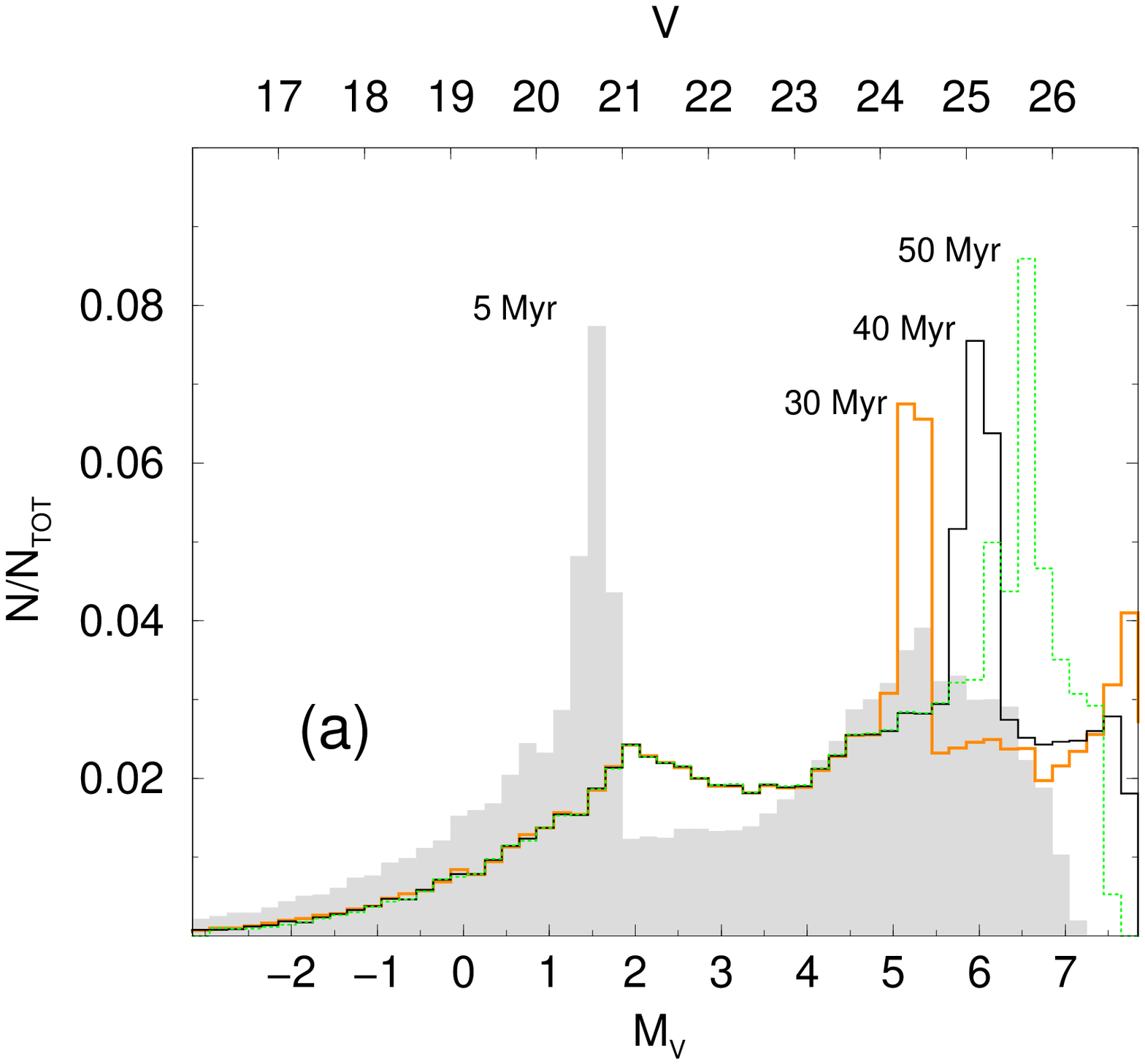}
\centering\includegraphics[width=7cm]{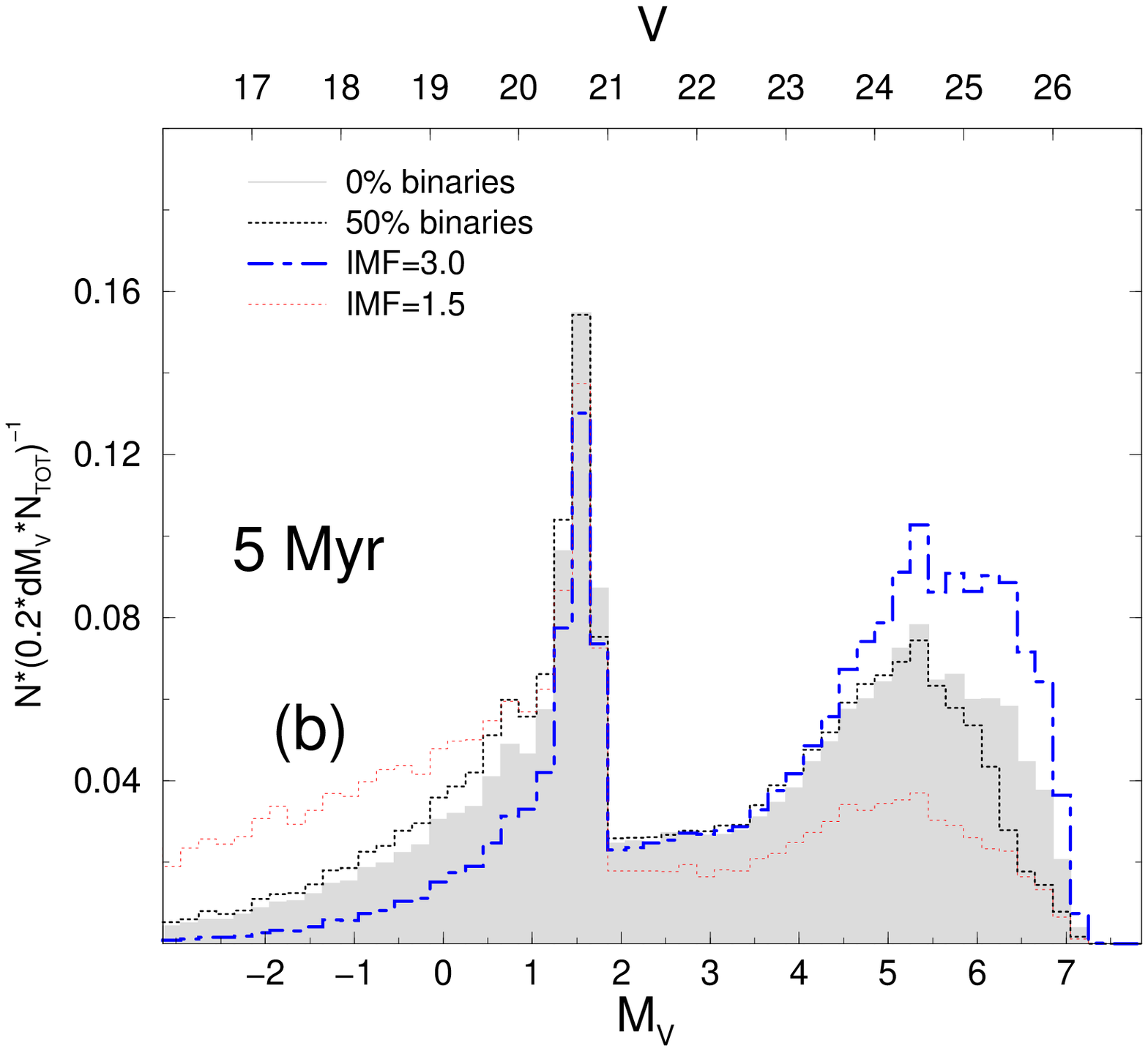}
\centering\includegraphics[width=7cm]{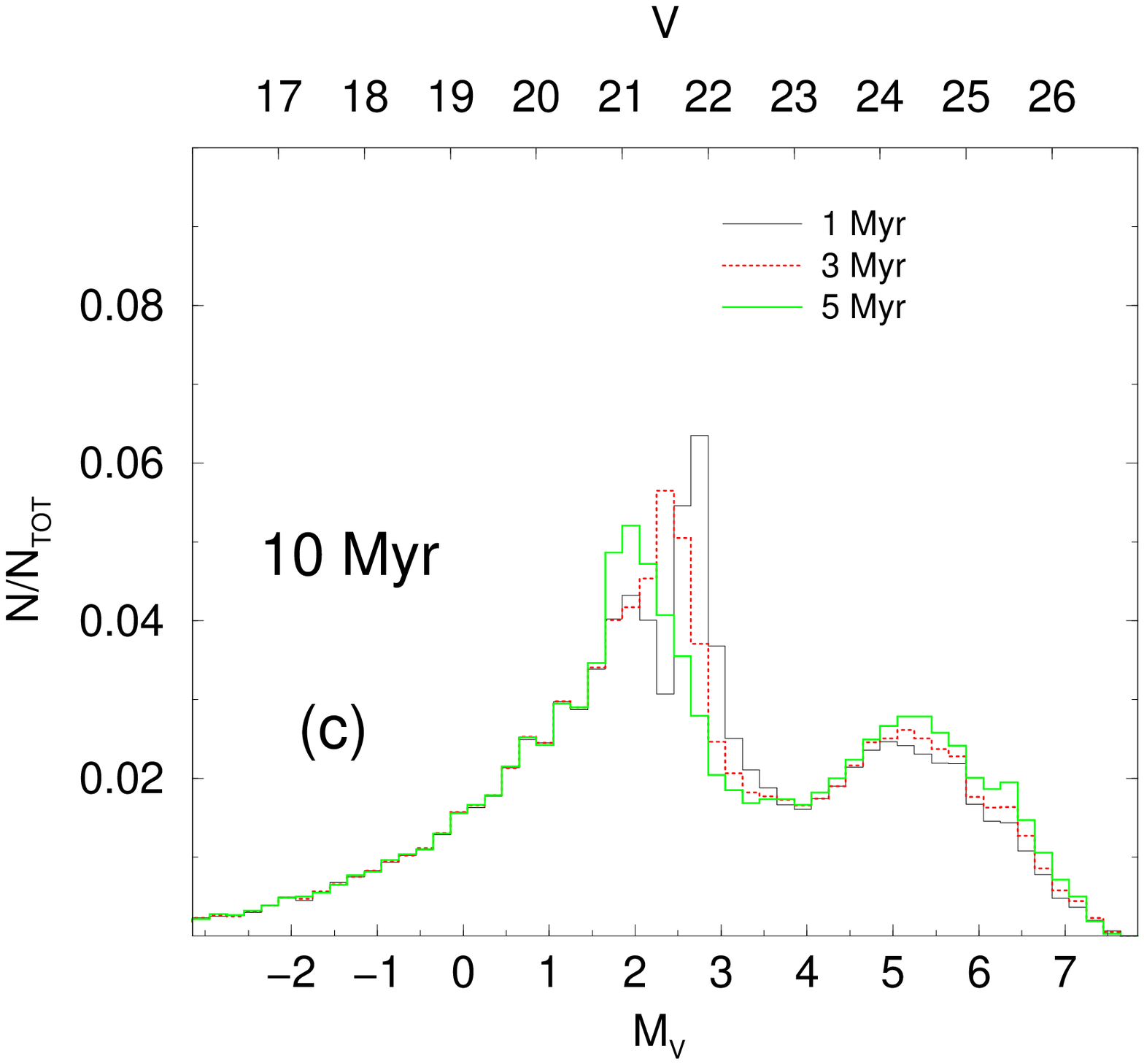}
\caption{Panel (a): Effect of incompleteness and photometric errors on
  the LFs for the labeled ages. All LFs are normalized to the total
  number of stars. Panel (b): Effect of different IMFs and binary
  fractions on the 5 Myr LF. The shadowed histogram is the same as the
  reference 5 Myr case of panel (a). Panel (c): normalized LFs for a
  10 Myr cluster with the labeled duration of prolonged star formation
  activity.}
\label{uncer} 
\end{figure}

Although attractive, the TOn dating method should be used with caution. When we
compare the observed and theoretical LFs it is important to evaluate
the following uncertainties:

\emph{Incompleteness and photometric errors.} To test real conditions,
the synthetic SSPs have been degraded assuming reddening and distance
of the SMC, $E(B-V)=0.08$ and $(m-M)_0=18.9$, and photometric errors
and incompleteness as derived by \cite{Sabbi07} from HST images of
NGC346. Fig.\ref{uncer}(a) shows the degraded LF for populations of
5, 30, 40, 50 Myr: at the distance of SMC the excellent quality of the
HST/ACS photometry guarantees perfect detectability of the peak/dip
feature up to about 50 Myr. Beyond this age the TOn becomes too faint to 
be detected.

\emph{Poisson fluctuations.} In order to check how many stars are
necessary to safely identify the TOn, we progressively reduced the
number of synthetic stars belonging to the 5 Myr population, until the
TOn peak was hidden by the Poisson fluctuations. This experiment
indicates that about 50 stars brighter than $M_{V}\approx 5$
(corresponding to a cluster total mass of $\approx\,500\,M_{\odot}$)
are sufficient to identify the TOn with a significance of $2\,\sigma$.

\emph{IMF, binaries, star formation duration.} In Fig.\ref{uncer}(b)
the synthetic 5 Myr SSP has been modeled changing the binary fraction
(50\% of simulated stars have now an unresolved companion, randomly
extracted from the same IMF as the primary) and the IMF exponent
(1.5-3). None of these changes alters significantly the TOn position:
the case with binaries is almost identical to the reference case,
while the adoption of different IMFs modifies only the shape of the LF
before and after the TOn peak.

In Fig. \ref{uncer}(c) we have simulated a 10 Myr old cluster with a
star formation activity that lasts 1, 3 and 5 Myr (see
e.g. \citealt[][]{Palla00}). As expected, with a prolonged SF activity
the LF peak grows brighter and broader.  In this case, the peak
magnitude provides information on the average age of the population.

\emph{Reddening.}  Typical of star forming regions, the presence of
highly obscuring material (foreground as well as local) can
significantly dim and blur a TOn. Thus, reliable reddening estimates
are fundamental to obtain unbiased ages. On the other hand, there are
several regions with affordable foreground and intrinsic
extinction. NGC346 is one of them with foreground $E(B-V)\sim 0.08$
and internal $\lesssim\,0.1$ mag (as deduced from the upper main
sequence), which contributes to the final uncertainty by $\approx$ 1-2
Myr.

\begin{figure*}[!t]
\centering \includegraphics[width=5cm]{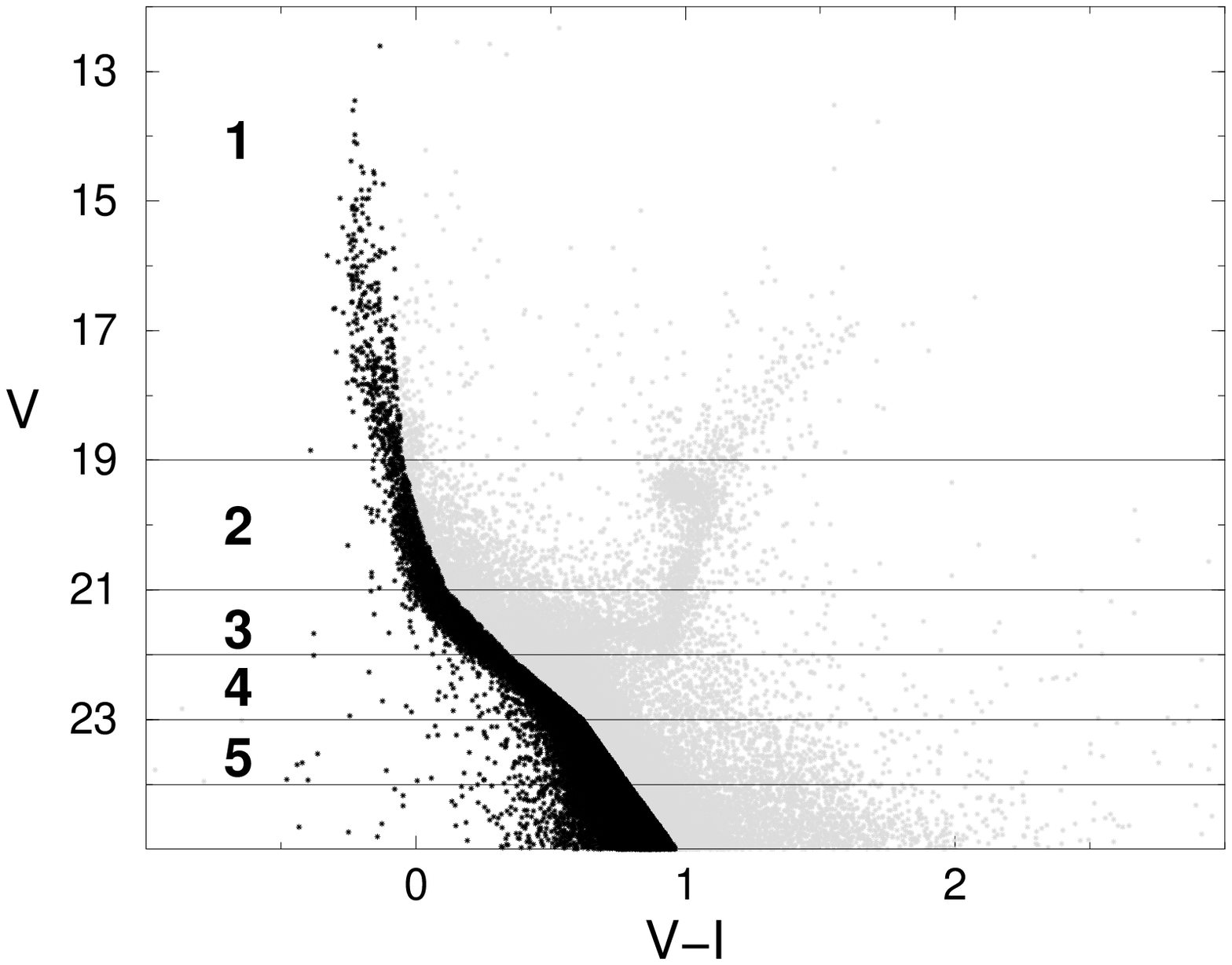} 
\centering\includegraphics[width=5cm]{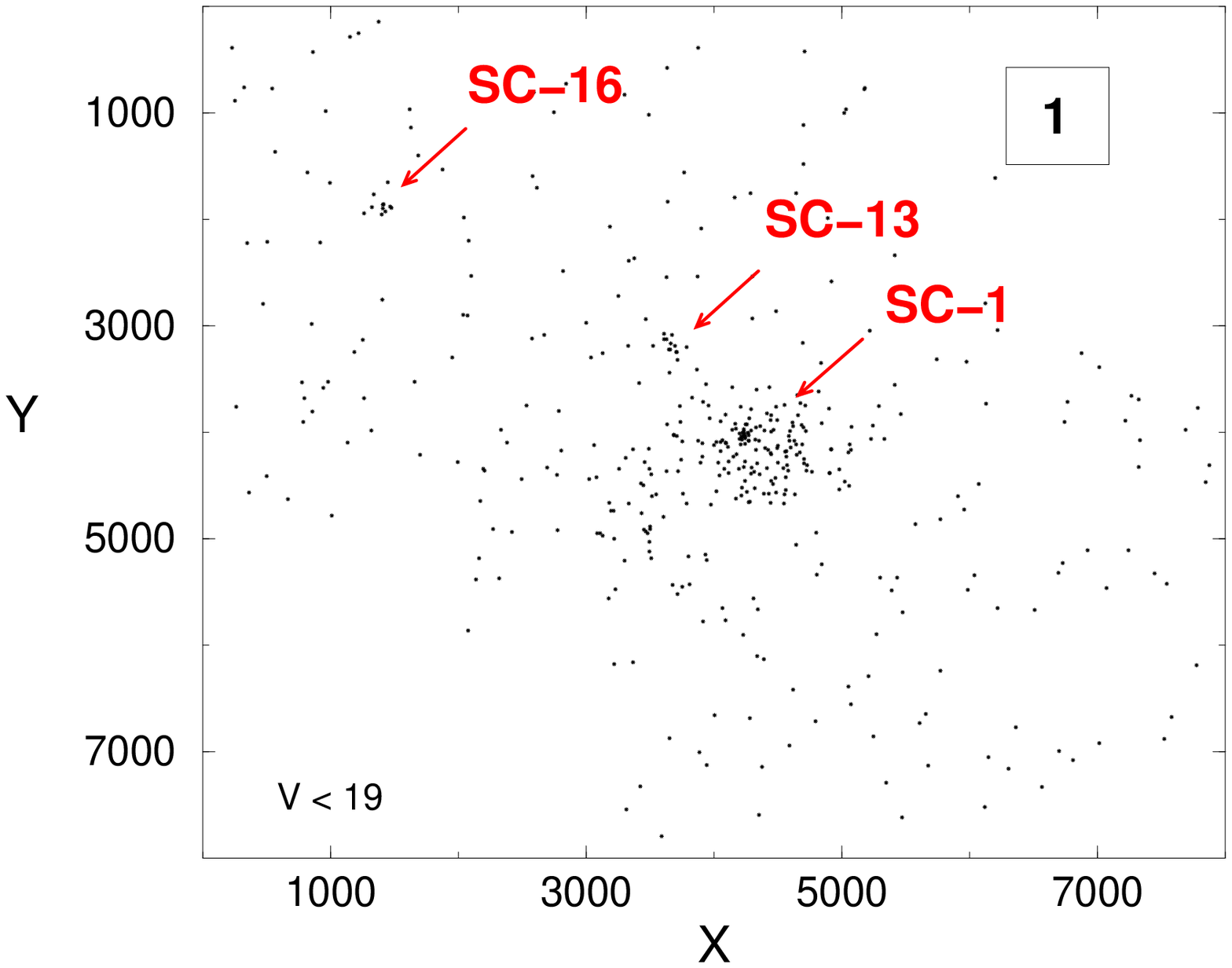}
\centering\includegraphics[width=5cm]{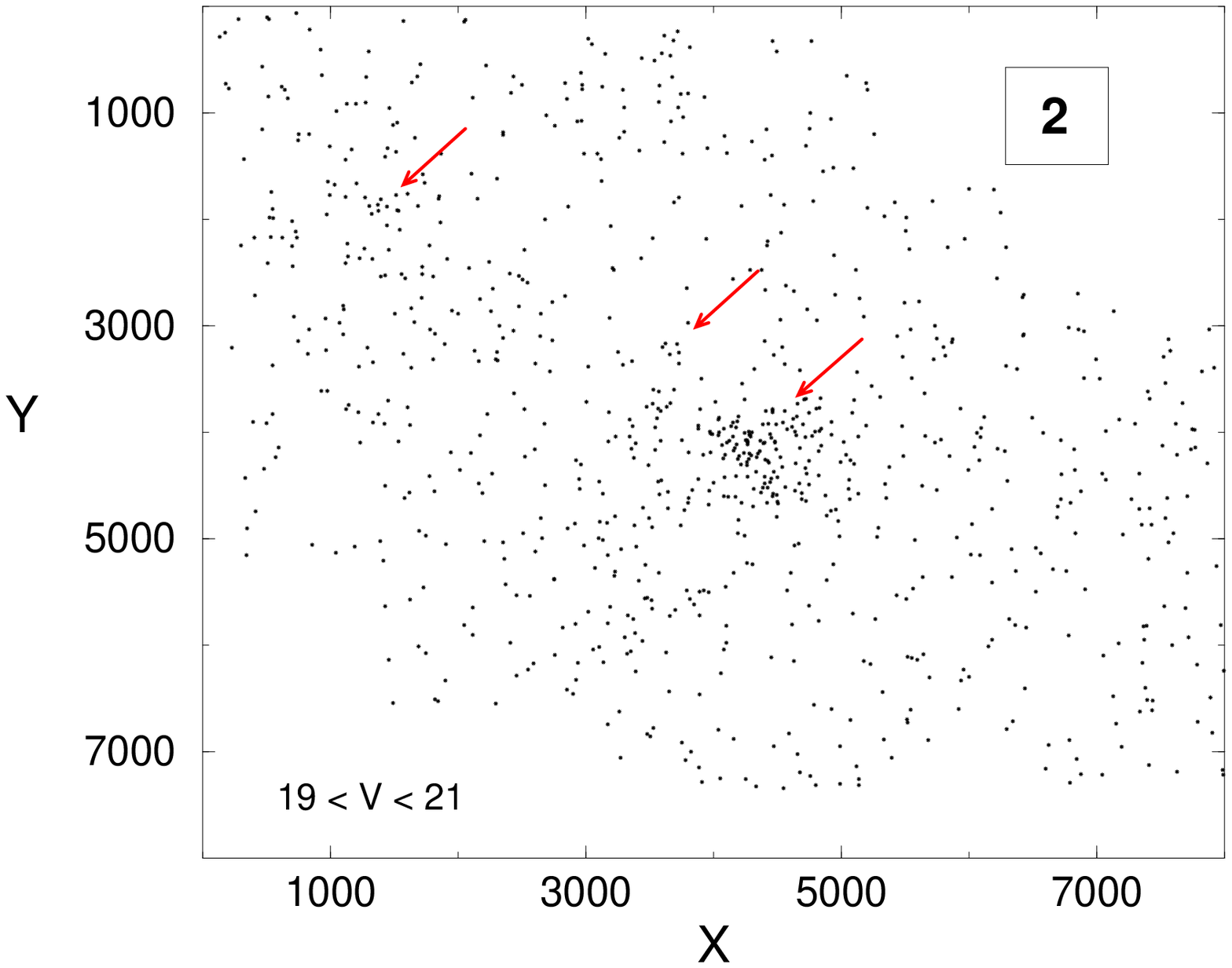}\\
\centering\includegraphics[width=5cm]{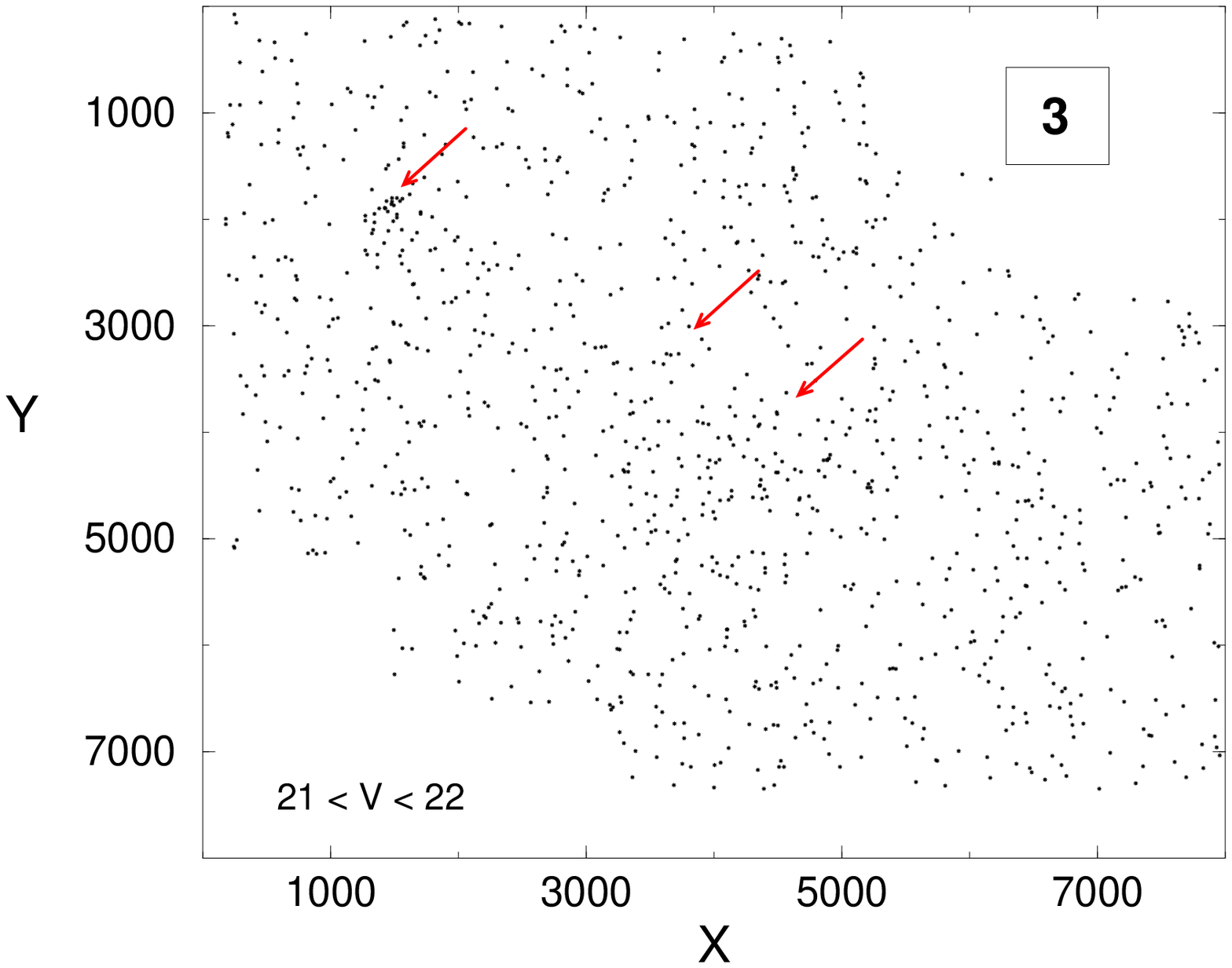}
\centering\includegraphics[width=5cm]{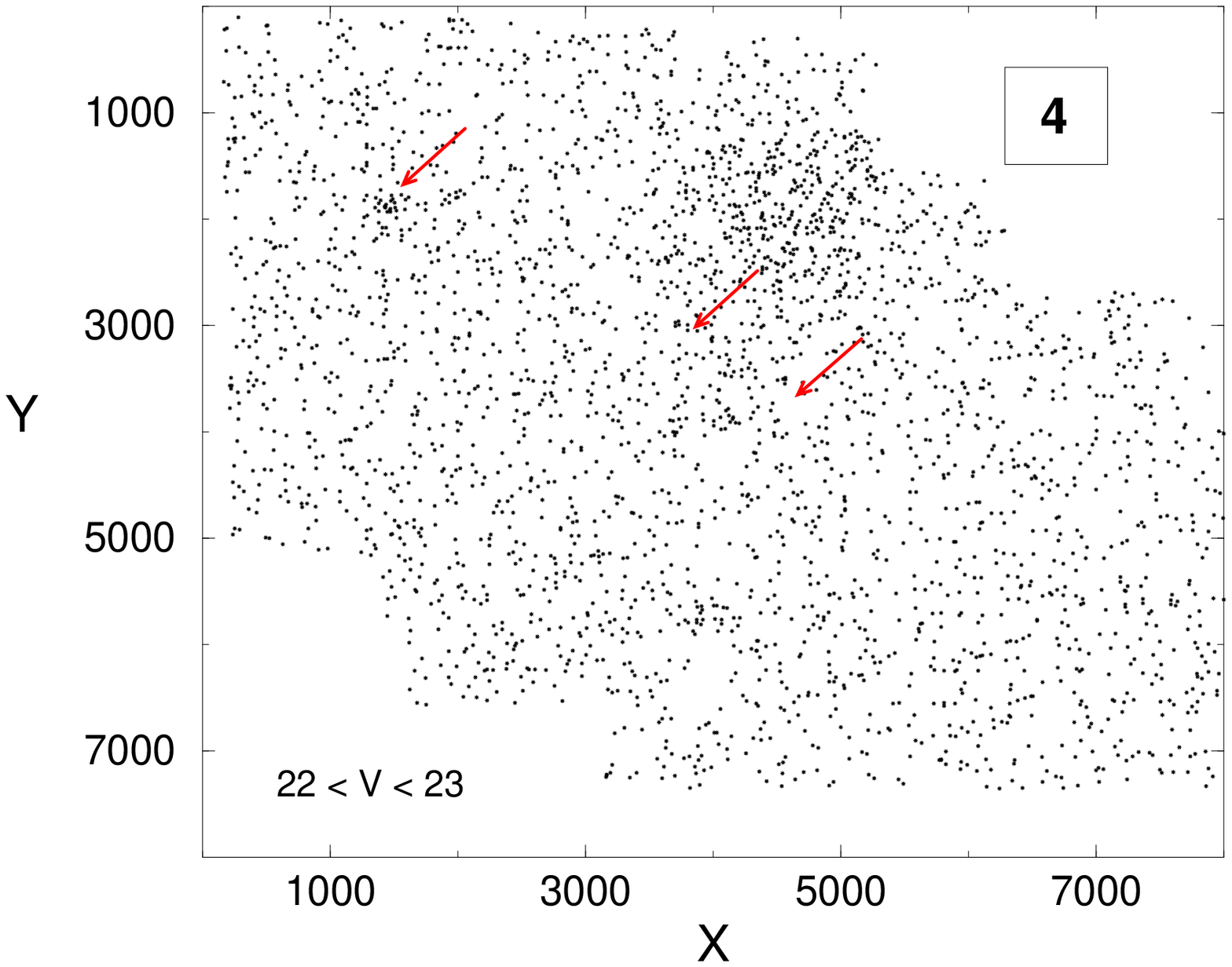}
\centering\includegraphics[width=5cm]{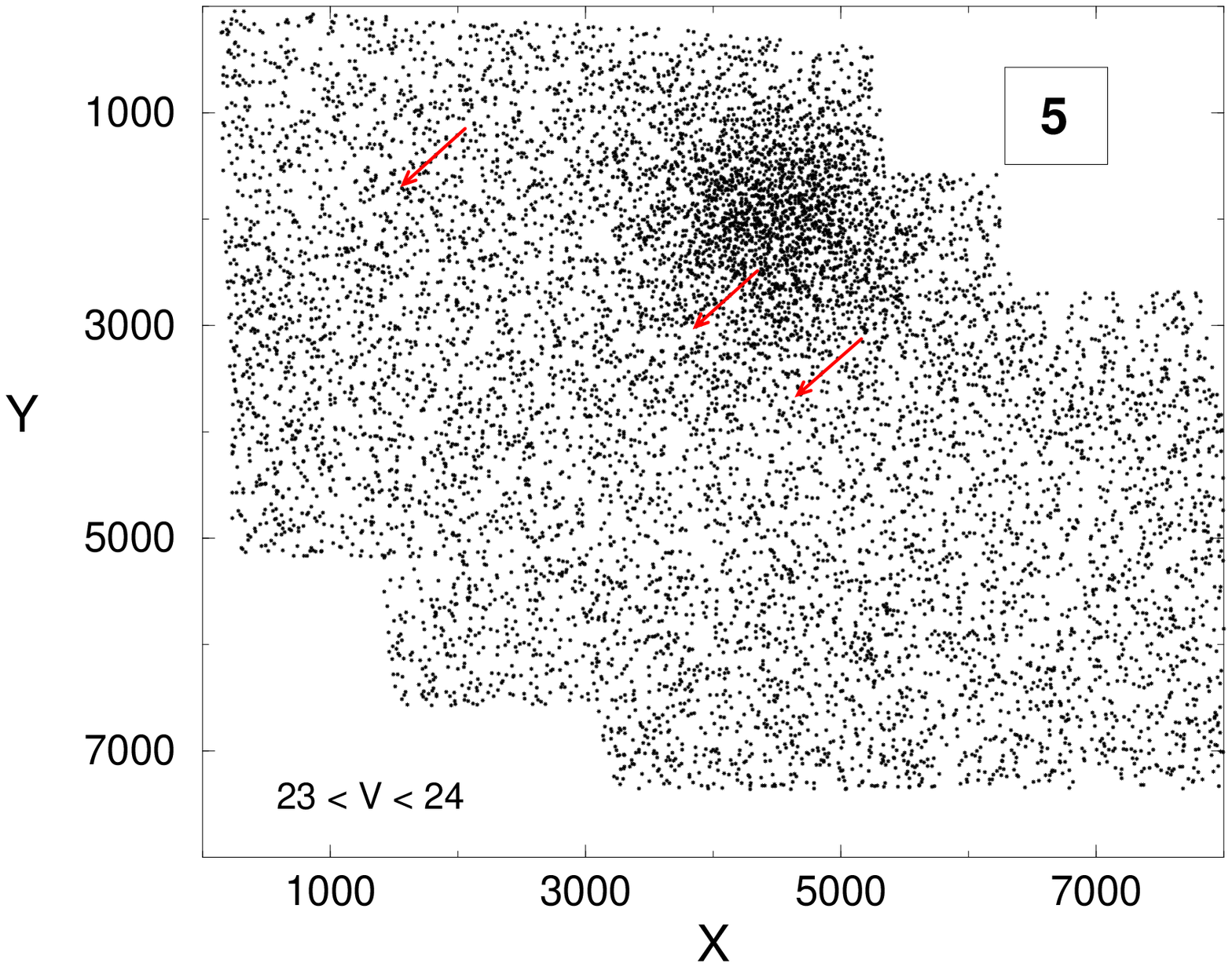}
\caption{The top-left  panel shows the CMD selection (see text) of bona-fide
  MS stars (black dots). Each of the other panels shows the spatial
  distribution of these MS stars for the labeled range of magnitudes. }
\label{ms_slices} 
\end{figure*}

\section{Identification of the TOn in extragalactic star forming regions: the case of NGC346}

Our goal is to study how star formation develops in extragalactic star
forming regions. In the CMD of these regions, the MS is often
contaminated by young fore/background stars and only in a few cases
membership information is available for a safe decontamination. The
presence of the field MS partially fills the LF dip, and sensibly
lowers the significance of the TOn in the LF.

To get around the problem of contamination we propose to combine the
fact that, by definition, no sub-cluster member on the MS can be
fainter than the TOn, \emph{together} with a careful analysis of the
spatial distribution of the stars in the region. The procedure we
suggest has 3 main steps: 1) Selection of bona-fide MS stars (both
members and non-members of the clusters) of all magnitudes. In order
to take into account photometric errors and reddening, we consider all
the stars bluer than a ridge line appropriately redder than the
theoretical ZAMS. 2) Division of the selected MS stars into bins of
progressively fainter magnitude. For each bin of a given magnitude,
the spatial distribution of stars is examined to assess whether the
sub-clusters are still visible. When a sub-cluster is clearly
identified up to a given magnitude, but it disappears in the fainter
maps, we identify the magnitude of its TOn.  This is because we have
reached the magnitude where the sub-cluster members have not yet
reached the MS, and MS stars fainter than this limit do not belong to
that sub-cluster. In order to evaluate the statistical significance of
the sub-cluster disappearance from the maps, for each of the three
sub-clusters we built a stellar density radial profile and we
evaluated the contribution from different magnitude bins. For field
stars we expect a flat stellar density profile, while a decrease in
stellar density is the typical signature of a sub-cluster. The range
of magnitudes where the transition between the two profiles occurs
identifies the apparent magnitude of the TOn. 3) Age determination of
the sub-clusters. After applying reddening and distance modulus
corrections, we use Eq. (\ref{eq1}) to translate the sub-cluster TOn
magnitude into an age. The final uncertainty is fixed by the bin
width, which is, in turn, chosen to obtain an acceptable number of
counts per bin.

To test the strength of our method, we applied it to the star forming
region NGC346 in the SMC, whose images acquired by the HST Advanced
Camera for Survey have revealed a wealth of young sub-clusters
containing several PMSs \citep{Sabbi07}. This region provides
excellent conditions to test our method because its recent strong
activity supplies an outstanding sample of PMS stars
\citep{Nota06}. The complexity of its structure, with several non
coeval sub-clusters, requires a method able to trace the temporal
sequence of events leading to the present configuration. In this
letter we focus on three of the richest sub-clusters, SC-1, SC-13 and
SC-16, as defined by \cite{Sabbi07}. Their location in the region can
be seen in their Fig.8.

Fig.\ref{ms_slices} summarizes our analysis of NGC346. To select only
MS stars we used a ridge line 0.05 mag redder than the theoretical
$Z=0.004$ ZAMS. We considered all the stars bluer than this line as
bona-fide MS stars, and divided them in six bins of different
size. Bona-fide MS stars are drawn in black in the top-left panel of
Fig.\ref{ms_slices}, where all the other stars are marked in grey. The
other five panels show the spatial distribution of the bona-fide MS
stars for progressively fainter bins of magnitude. Analyzing these
maps, we find that the central sub-cluster SC-1 is well visible down
to $V=21$, while no obvious spatial structure resembling SC-1 is
observable in the fainter maps. The sub-cluster SC-16, on the
contrary, is still clearly visible down to $V=23$. Beyond this limit,
also SC-16 vanishes. The small sub-cluster SC-13 is recognizable at
least to $V=21$. Notice that the stellar agglomeration appearing at
$V=22$ corresponds to the 4-5 Gyr turn-off stars of the older cluster
BS90 (see e.g. \citealt{Sabbi07}) in the foreground of NGC346.

In order to test the significance of the sub-clusters we examined the
radial density profile of the three sub-clusters as a function of
magnitude (Fig. \ref{res} left panels). The profiles are calculated
using annuli of equal area centered on the highest density peak. The
inner radius is fixed at 150 pixel. To exclude any completeness issue
in these crowded regions, we also show in the middle panels of Figure
\ref{res} the completeness factors obtained from the extensive
artificial star tests performed by \cite{Sabbi07}, for all stars
within 100 pixels from the centers of SC-16, SC-1 and SC-13. To
further test the final ages, theoretical isochrones are over-imposed
on the CMDs (right panels of Fig.\ref{res}).

\subsection{SC-16}
The first row in Fig.\ref{res} shows the MS radial profiles (left
panel), the completeness curves (middle panel) and the CMD (right
panel) for SC-16. The compact morphology of this sub-cluster is
evident: the radial profile of MS stars brighter than $V=23$ rapidly
drops with distance from the center, while the distribution of less
luminous MS stars is constant. This confirms that the TOn magnitude is
between 22 and 23, thereby constraining the first generation of stars
to have formed between 12.5 and 18 Myr ago.  The upper right panel of
Fig.\ref{res} shows the corresponding isochrones superimposed to the
SC-16 CMD (all stars within 150 pixel from the center): it is
reassuring to see that below the 18 Myr TOn, the sub-cluster stars
move away from the MS, confirming that the MS deficit in the map is a
genuine evolutionary effect. Concerning the completeness, although the
sub-cluster region loses stars faster than the field, for magnitudes
$V<23$ this effect is not severe (compared to the field, less than
25\% of stars are lost). By comparing our age estimate to independent
determinations we find good agreement both with \cite{Sabbi07} ($15\pm
2.5$ Myr) and with \cite{Hennekemper08} (5-15 Myr).

\subsection{SC-1}
 The case of SC-1 (second row in Fig. \ref{res}) is quite
 different. Here the extended spatial morphology produces a gradual
 decrease in the MS radial profile rather than an abrupt drop at
 $V\approx 21$. Radial profiles for fainter magnitudes are quite
 flat. Assuming a conservative bin uncertainty of 1.5 mag on the TOn
 magnitude, the star formation onset is expected between 3.5 and 6.5
 Myr ago, in agreement with the estimate by \cite{Sabbi07} ($3\pm 1$
 Myr). Both the 3.5 and the 6.5 isochrones are overlaid on the SC-1
 CMD. As suggested by radial profiles at $V=21$ there is a clear
 signature of a PMS TOn, with a large sample of stars still in the PMS
 phase. Our 6.5 Myr isochrone fits the upper MS morphology, including
 the TOn, very well and fits the PMS blue envelope as well. The few MS
 stars below the TOn are likely contamination of the SMC field and/or
 stars of the old cluster BS90 (the presence of 3 clump stars endorses
 this hypothesis). The absence of any star on the MS around $V\sim
 21.5$ and the morphology of the CMD for the vast majority of the
 stars, which seems to follow the isochrone with PMS included, further
 supports our conclusion. The measured completeness (middle panel) is
 better than 75\% and cannot account for such a reduction.

\subsection{SC-13}
\begin{figure*}[!t]
\centering \includegraphics[width=5cm]{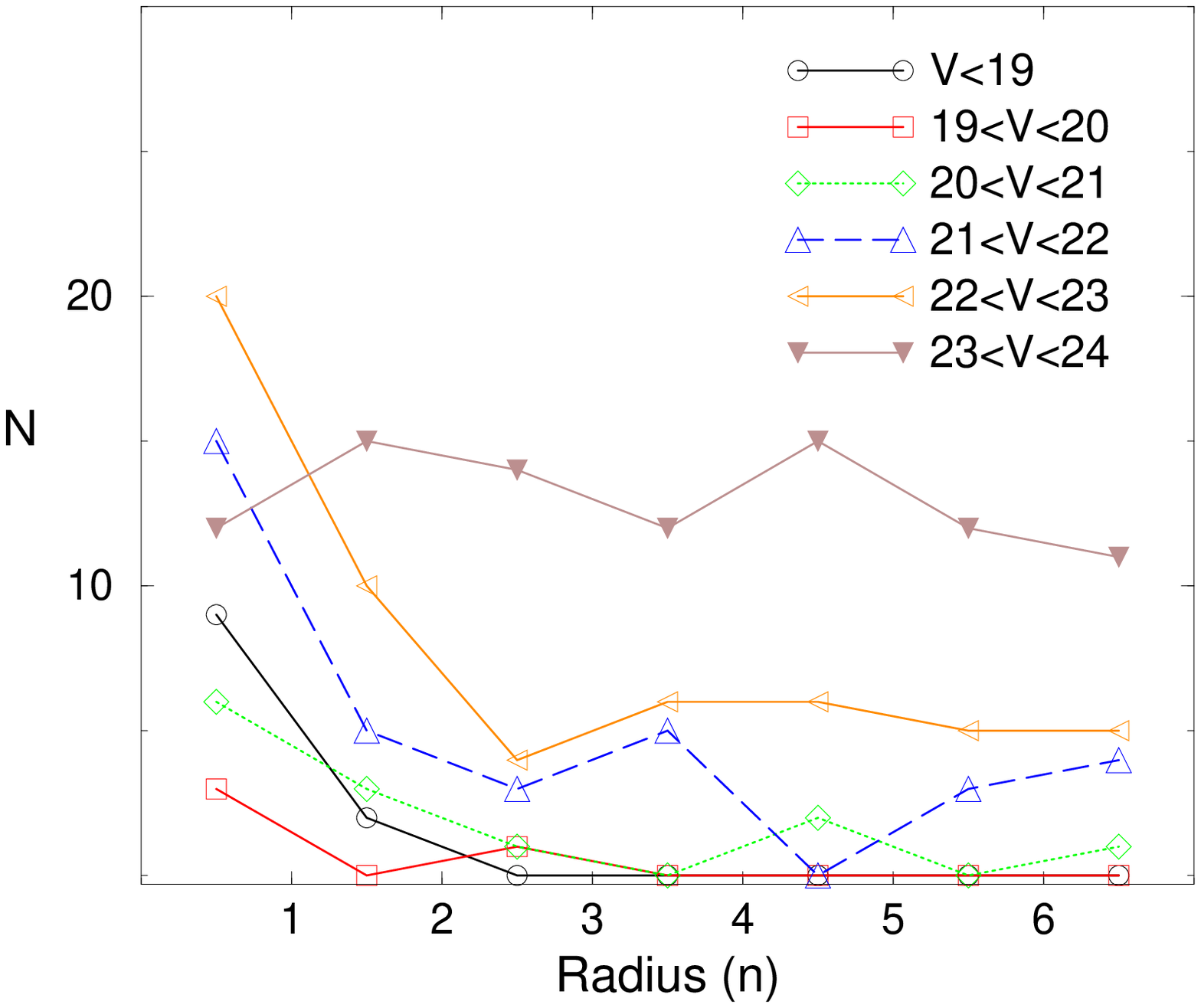}
\centering \includegraphics[width=5cm]{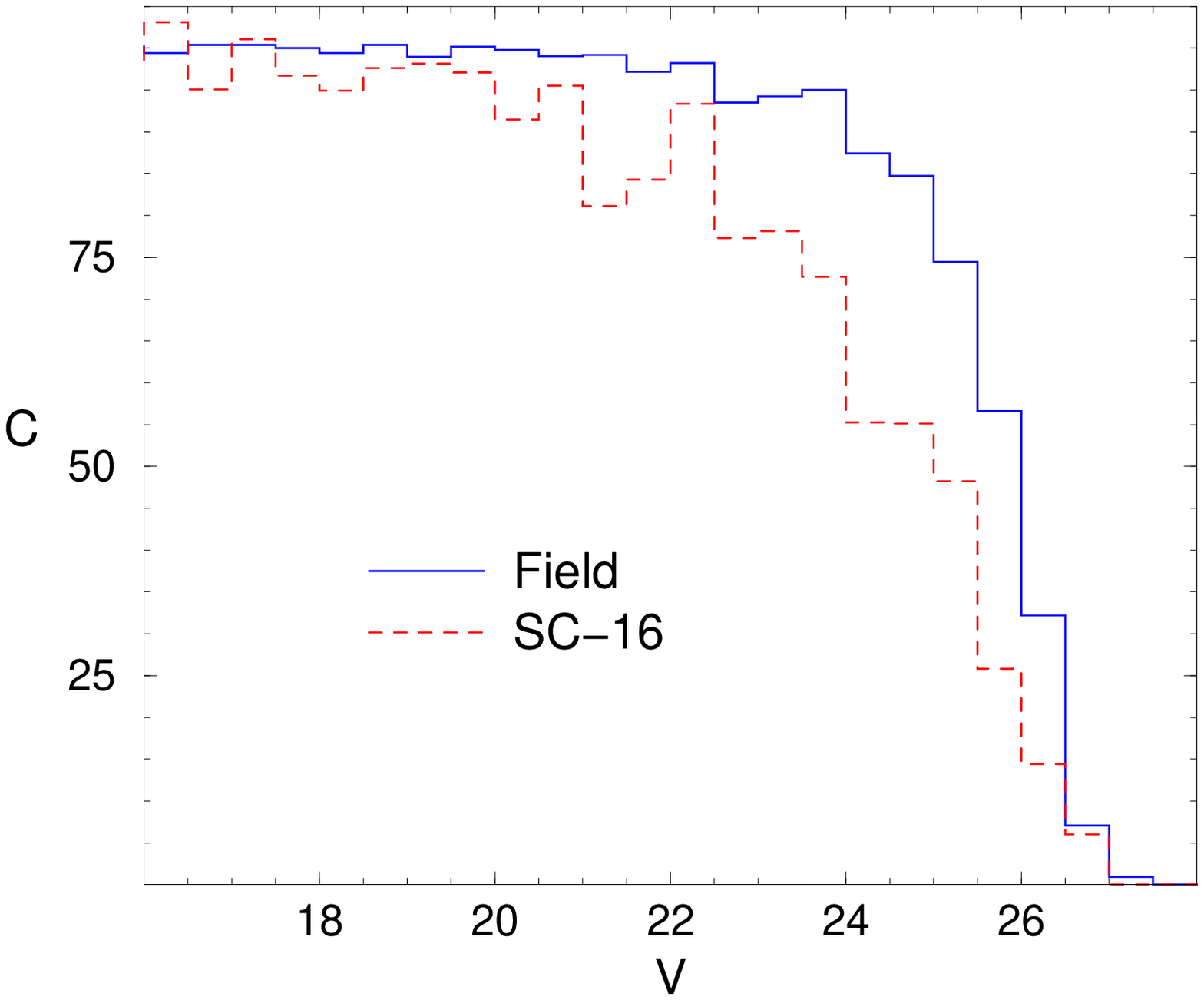}
\centering \includegraphics[width=5cm]{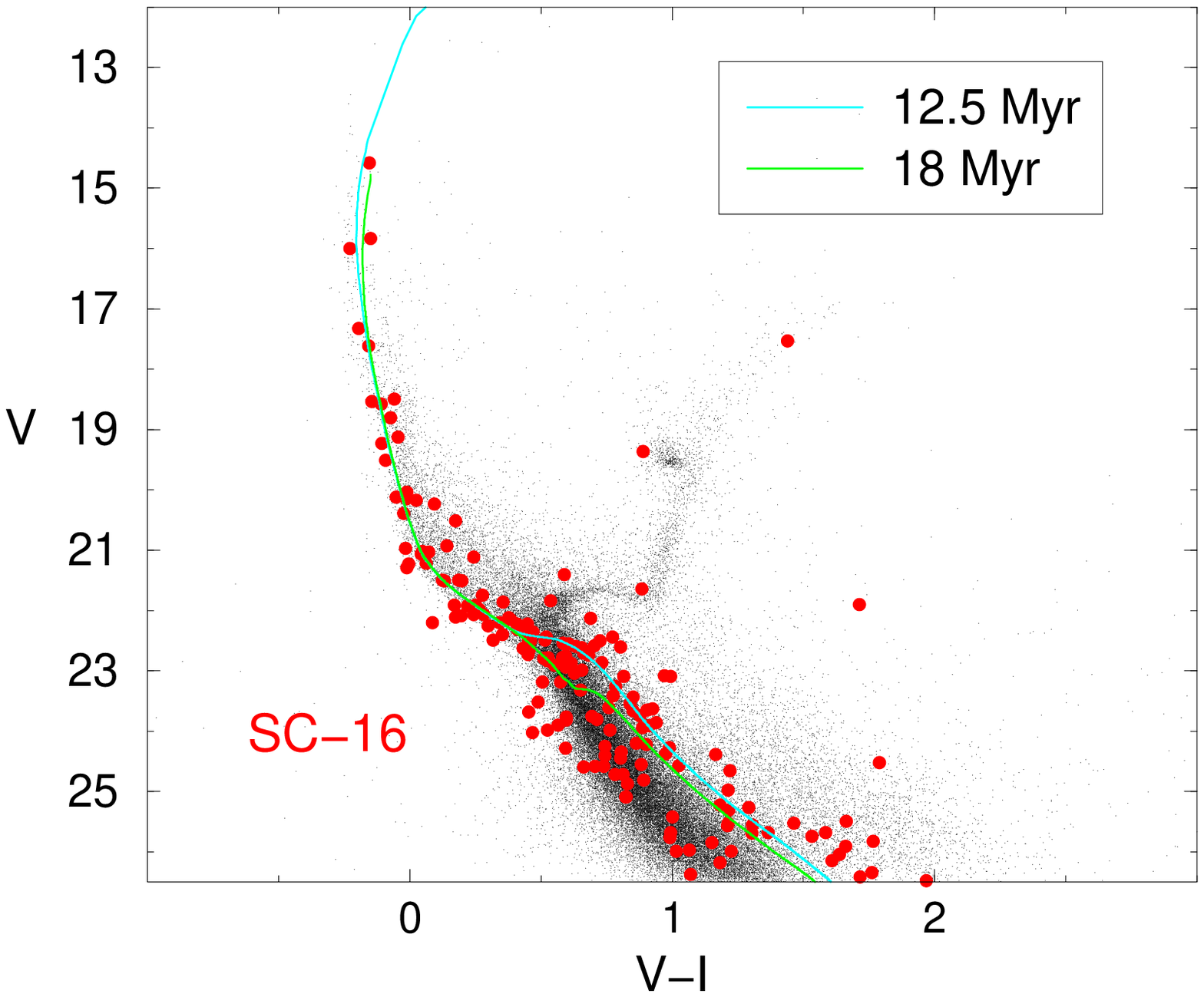}\\
\centering \includegraphics[width=5cm]{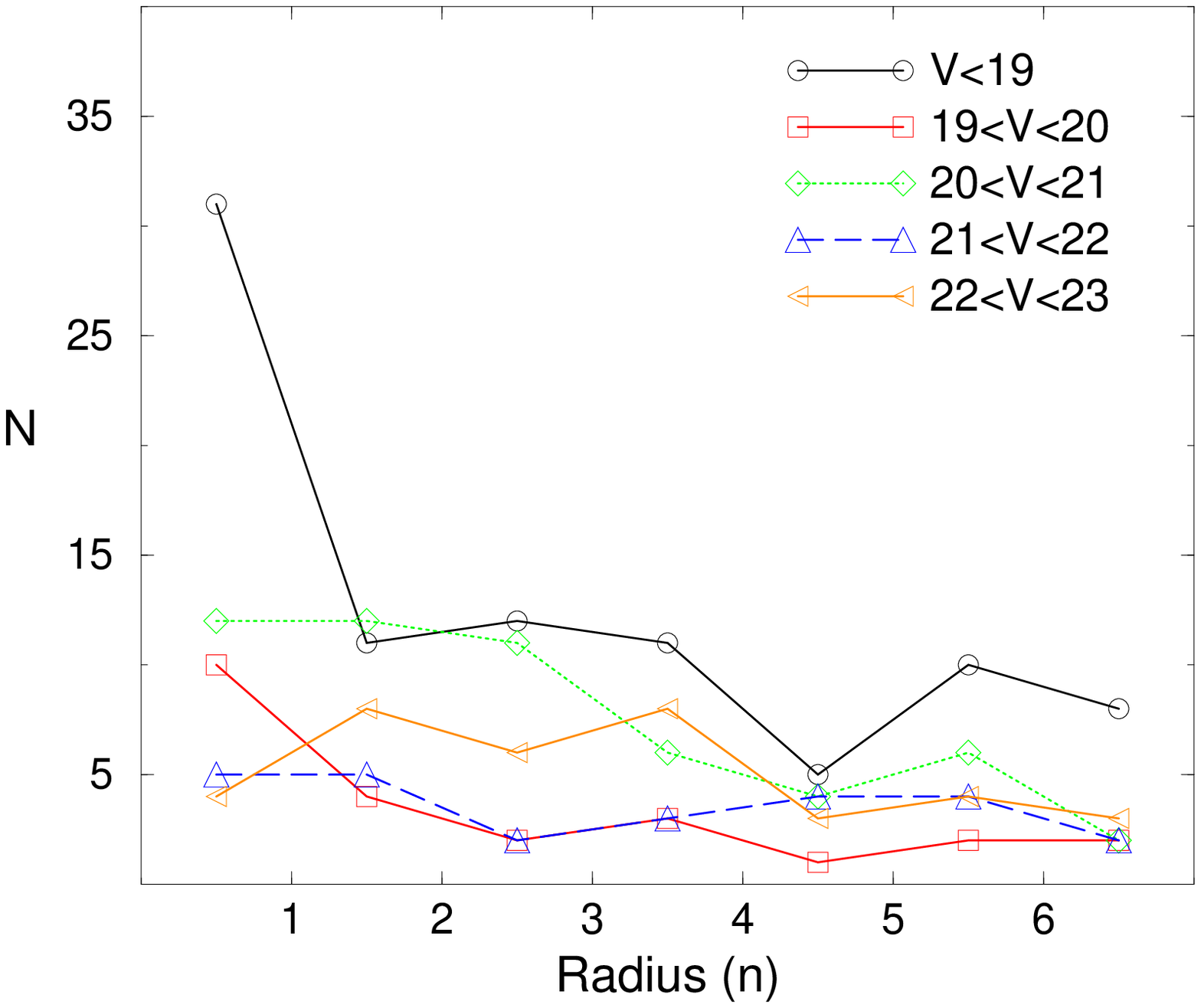}
\centering \includegraphics[width=5cm]{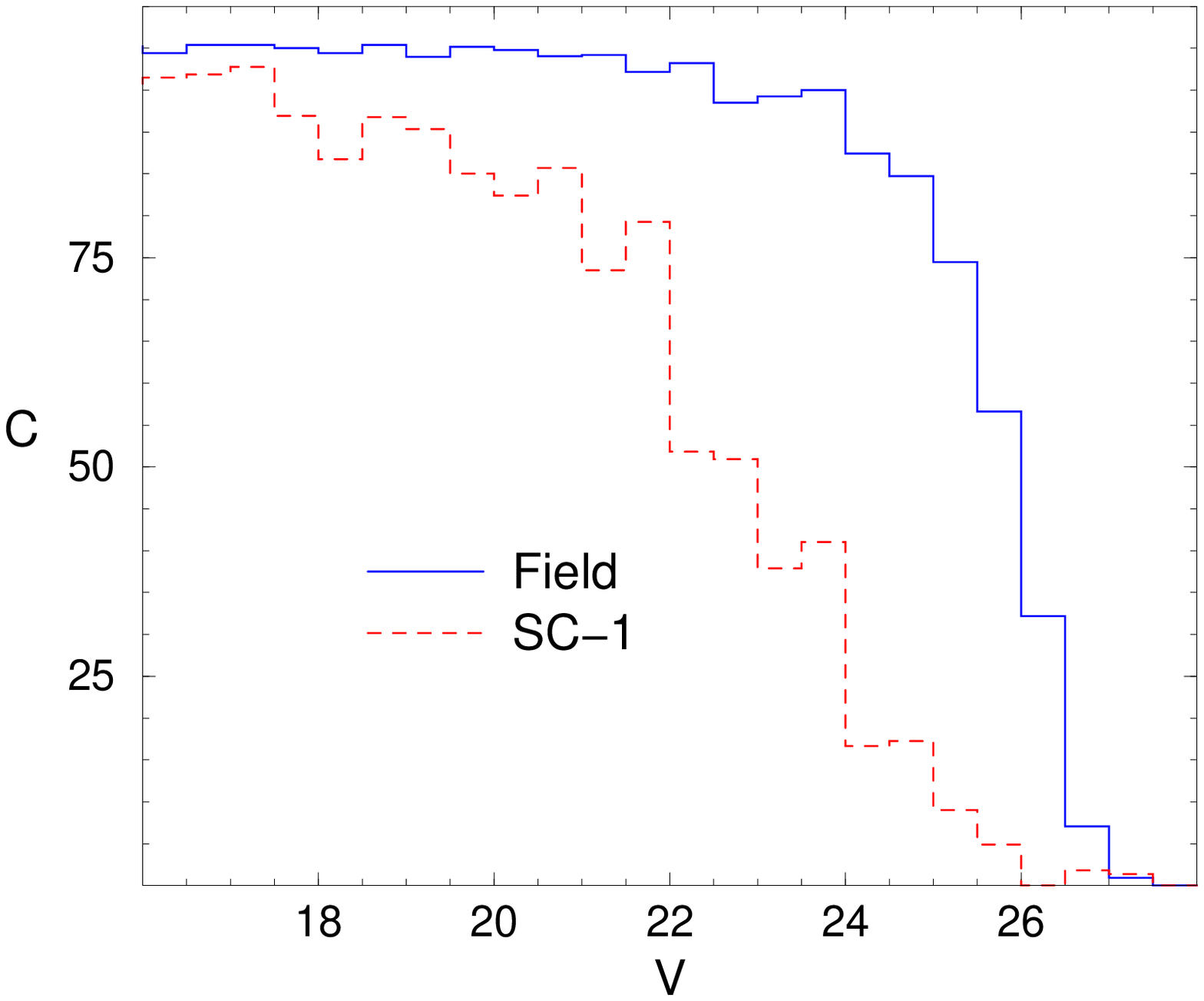}
\centering \includegraphics[width=5cm]{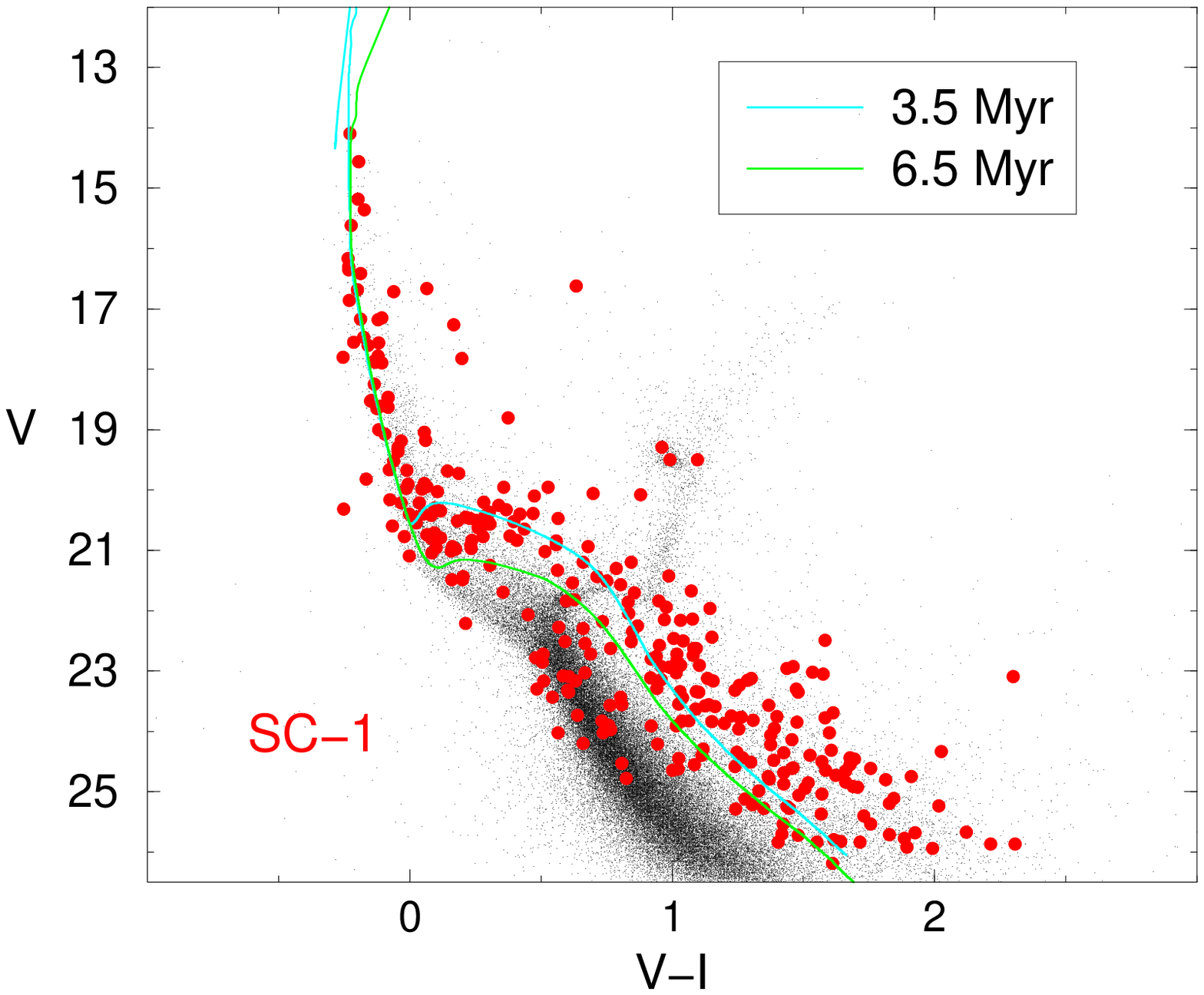}\\
\centering \includegraphics[width=5cm]{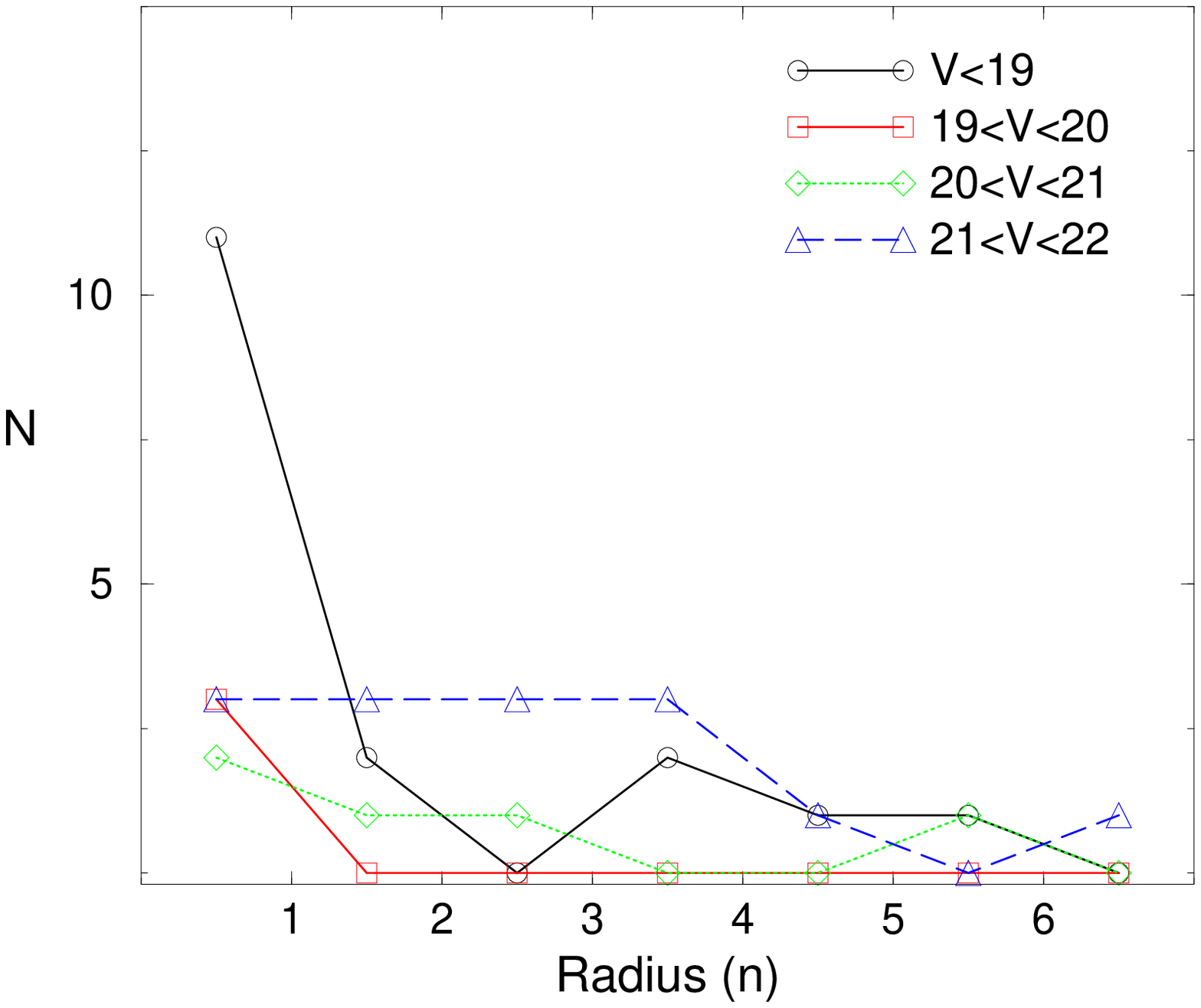}
\centering \includegraphics[width=5cm]{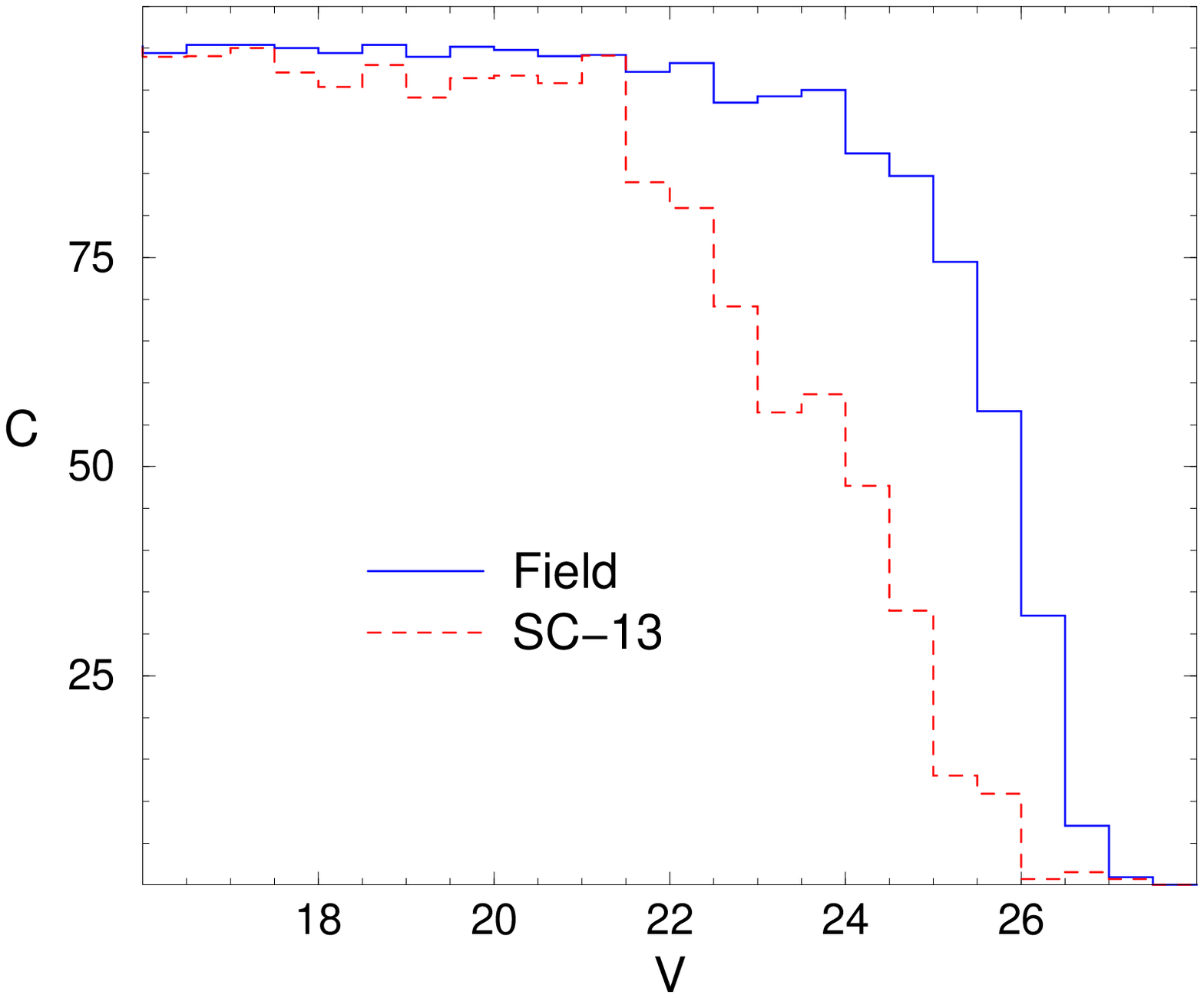}
\centering \includegraphics[width=5cm]{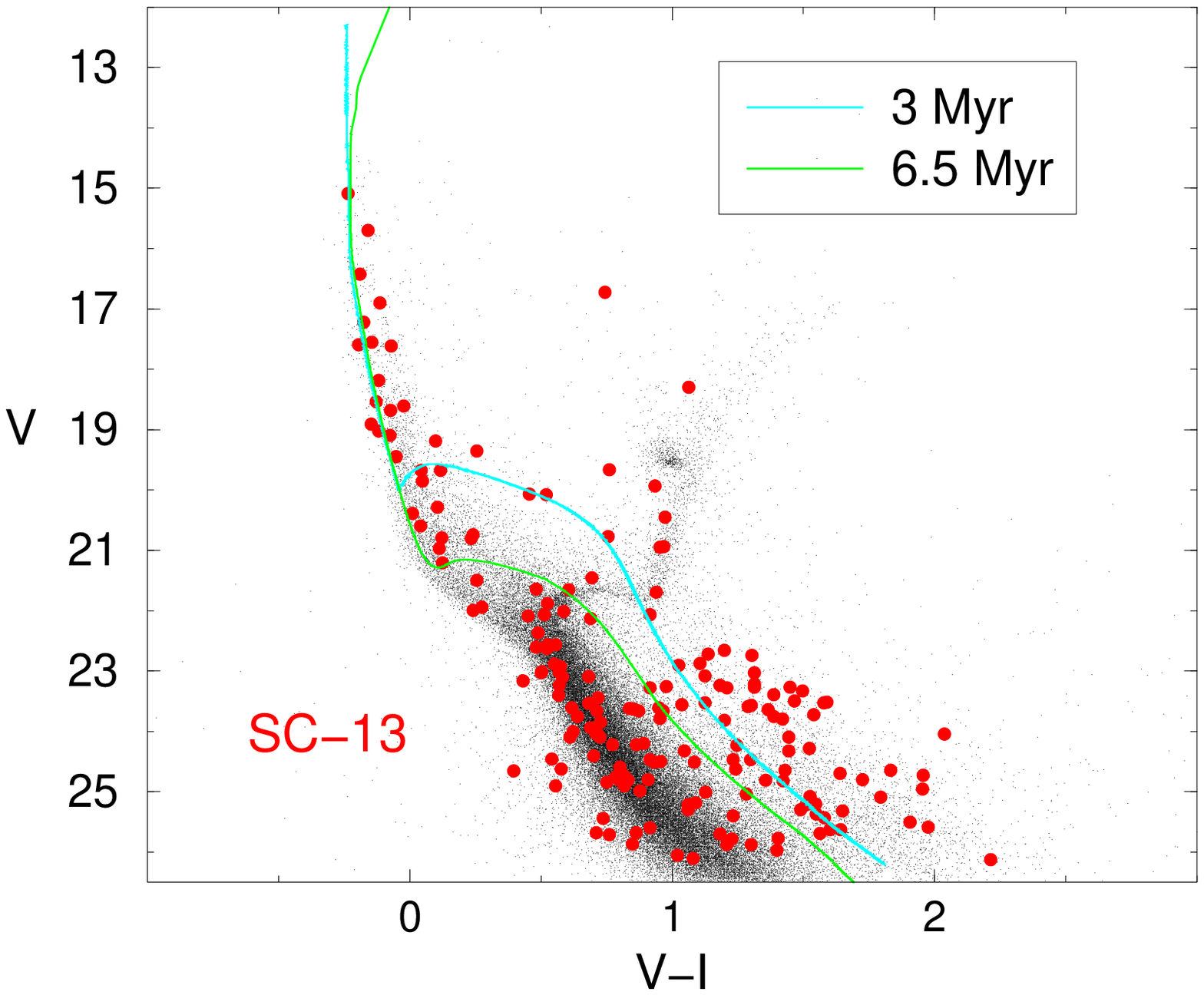}
\caption{Left-hand panels: radial distribution of MS stars near the
  centers of SC-16 (top), SC-1 (middle) and SC-13 (bottom). Different
  curves correspond to the indicated ranges of magnitude. The abscissa
  is in units of $n$, where $n$ is related to the distance $d$ from
  the sub-cluster center by
  $d=\sqrt{n}\times\,150\,\mathrm{pixels}$. Middle panels:
  Completeness for the labeled sub-clusters (dashed line) and field
  stars (solid line). Right-hand panels: CMDs for all stars within 150
  pixels from the center of SC-1 (top panel), SC-16 (middle panel) and
  SC-13 (bottom panel).  Red dots indicate sub-cluster stars while
  black dots stand for the entire data sample. Isochrones of the
  indicated ages are also shown.  }
\label{res} 
\end{figure*}

The third row of Fig. \ref{res} shows the sub-cluster SC-13. As for
SC-1, the MS density profiles of SC-13 are declining down to
$V>21$. However, due to the poor statistics, we can only state that
the cluster formation started sometimes in the last 6 Myr. There is a
further clue suggesting that SC-13 is actually younger than 6 Myr: a
close inspection of the CMD (right panel) reveals that most of
probable intermediate-mass PMS stars are aligned with the 3 Myr
isochrone. In agreement with this finding, most of the low-mass PMS
stars are redder than the 3 Myr isochrone. This age is also found by
\cite{Sabbi07} ($3\pm 1$ Myr) and by \cite{Hennekemper08} (0.5-2.5
Myr).

\section{Discussion and Conclusions}

The evolution of young star forming regions in their earliest stages
is still quite unknown. For very young clusters, with ages of tenths
to few Myr, the identification of the turn-off is usually hampered by
the paucity of massive stars. However when a cluster is sufficiently
young to harbor PMS stars, the luminosity of its TOn provides a robust
indication of its age.  Furthermore, while the PMS
evolutionary models still suffer of several limitations, the simple
comparison of TOn luminosities is a reliable measure of relative ages.

In the LF the TOn is a narrow peak followed by a dip. The TOn is very
sensitive to age: for a $\sim 30$ Myr old stellar population the
luminosity of the TOn changes by $\sim 0.08$ mag/Myr, but at $\sim 20$
Myr it already changes by $\sim 0.15$ mag/Myr, and by $\sim 0.33$
mag/Myr at 3 Myr.

To guarantee a safe TOn identification, it is important to select
targets with low or well known reddening. In the visual, this
currently excludes Galactic clusters like Westerlund 1-2, NGC3603,
Arches and Quintuplet, but good targets can be found in the solar
vicinity and in some nearby irregulars galaxies like the Magellanic
Clouds and IC1613.

In this letter we have presented a new method to investigate how star
formation develops in complex extragalactic young clusters such as
NGC346 in the SMC.

Our method combines the analysis of the star clusters stellar
densities profiles with the notion that in a cluster no star on the MS
can be fainter than the cluster TOn. This approach has the advantage
of strongly reducing the uncertainties introduced by the contamination
to the CMD by young stars that do not belong to the cluster,
ultimately affecting the TOn detectability in both CMD and LF.

Clearly, there are limitations to the applicability of the method. The
bona-fide MS selection may include intruders like PMS and stars
evolved off the MS, reducing the TOn visibility. The issue is
particularly thorny at ages larger than about 30 Myr, because the PMS
isochrones are close to the ZAMS (see Figure \ref{peaks}(a)). The
method relies on the assumption that the formation sites are
agglomerations of stars: if for some reason the newborn stars formed
in isolation or, drifting apart, were too diluted to be detected as
aggregates, the method would not be applicable. In this respect, the
study by \citet[][]{Pfalzner09}, exploiting the density-radius
relation to date nearby clusters younger than 20 Myr is interesting
and reassuring.

We applied the TOn method to NGC346 in SMC, and we found that the
onset of the star formation in the sub-cluster SC-1 was between
$3.5-6.5$ Myr, in the sub-cluster SC-13 3 Myr ago or less and in SC-16
between $12.5-18$ Myr ago, in good agreement with the results from the
literature.

Having established the effectiveness of our method, the next steps
will be to: 1) incorporate near-infrared photometry and use it to
study extinguished regions where measurements of the upper MS are
difficult to carry out; 2) undertake comparative studies of the
duration and spatial patterns of SF in Magellanic Cloud regions
ranging from the giant 30 Dor complex to the comparatively isolated
NGC602 cluster.

\section*{Acknowledgments}
We thank A. Bragaglia and S.N. Shore for useful suggestions. MC and MT
acknowledge financial support through contracts ASI-INAF-I/016/07/0
and PRIN-MIUR-2007JJC53X-001. Partial support for U.S. research in
program GO10248 was provided by NASA through a grant from the STScI,
which is operated by the AURA, Inc., under NASA contract NAS5-26555.

%\bibliography{myrefs}

\end{document}